\documentclass[12pt]{article}
\usepackage{amsbsy}
\usepackage{amsfonts}
\usepackage{amsmath}
\usepackage{amssymb}
\usepackage{apacite}
\usepackage{setspace}
\usepackage{graphicx}
\usepackage{bm,multicol}
\usepackage[english]{babel}
\usepackage{natbib}
\usepackage[T1]{fontenc}
\usepackage[utf8]{inputenc}
\usepackage{authblk}
\usepackage{xcolor}
\newtheorem{lemma}{Lemma}

\newcommand{\newc}{\newcommand}
\newc{\N}{\mbox{N}}
\newc{\1}{\bf{1}}

\begin{document}
\title{Bayes factor testing of equality and order constraints on measures of association in social research}

%\author{Joris Mulder \& John P.T.M. Gelissen}
\author{Joris Mulder$^{*\dagger}$ \& John P.T.M. Gelissen$^*$}
\date{\small
$^*$Department of Methodology and Statistics, Tilburg University, The Netherlands\\
Jheroniums Academy of Data Science, ’s-Hertogenbosch, The Netherlands
}

\maketitle

\begin{abstract}
Measures of association play a central role in the social sciences to quantify the strength of a linear relationship between the variables of interest. In many applications researchers can translate scientific expectations to hypotheses with equality and/or order constraints on these measures of association. In this paper a Bayes factor test is proposed for testing multiple hypotheses with constraints on the measures of association between ordinal and/or continuous variables, possibly after correcting for certain covariates. This test can be used to obtain a direct answer to the research question how much evidence there is in the data for a social science theory relative to competing theories. The accompanying software package `{\tt BCT}' allows users to apply the methodology in an easy manner. An empirical application from leisure studies about the associations between life, leisure and relationship satisfaction and an application about the differences about egalitarian justice beliefs across countries are used to illustrate the methodology.
\end{abstract}

\section{Introduction}
In the social sciences, statistical measures that quantify the degree of association between the variables under study are a fundamental tool for making inferences when it is not possible to assess the direction of the causal effects of interest. This may be the case for certain research designs or when substantive knowledge about the direction of the effects is missing. For example, sociologists who only have cross-sectional (observational) data at their disposal often use regression effects to test their substantive hypotheses although from a research design perspective other measures of association would be more appropriate. By testing regression effects they make strong, possibly questionable, assumptions concerning the causal direction of the variables that are involved in the analysis. Another example is when there is no substantive motivation for specific causal directions of the dependent focal variables under study. A major sociological field in which this occurs is sociological values and attitudes research where attitudinal dimensions are frequently non-causally related to each other. Also in this situation, testing hypotheses about associations between variables necessitates using measures of association other than regression coefficients.

%{\color{red}In the social sciences, statistical measures that quantify the degree of association between variables of interest are a fundamental tool for making inferences from correlational data when the design or the logic of the study make it implausible that a variable X is a causal antecedent to a variable Y. For example, sociologists who have only cross-sectional (observational) data at their disposal may use regression approaches to test their substantive hypotheses. Thereby they make strong assumptions concerning the causal direction of the variables that are involved in the analysis. Although the analysis of regression effects is the prevailing statistical approach in contemporary social sciences for analyzing associations in such data, from a research design perspective using other measures of association for inference would be more appropriate. Another relevant research situation is when the focal variables in the study can be assumed to be only associated with each other without any substantive reason to expect a specific causal ordering between them. A major sociological field in which this often occurs is sociological values and attitudes research where attitudinal dimensions are frequently non-causally related to each other. Also in this instance, testing hypotheses about associations between variables necessitates using measures of association other than the regression coefficient.}

The best-known measure of association is Pearson's correlation coefficient, which expresses the strength of the linear relationship between two continuous variables in the data. If the variables are measured on an ordinal (Likert-type) scale, ordinal measures of association such as Spearman's rho are needed to quantify the strength of the linear relationship between the variables. In many analyses, the researcher is not only interested in a zero-order association between the variables of interest, but the researcher also wants to rule out that any association found is the result of the two focal variables having a common cause, which would make the zero-order correlation spurious. A classical tool for this purpose is the partial correlation coefficient, which can be used to measure the linear association between two variables while controlling for other variables.

Many researchers usually stop after the second step in which other variables are controlled or partialled out, just to conclude whether there is enough evidence to conclude that a spurious association exists. But such a basic correlational analysis also has the potential for testing more complex and interesting substantive sociological hypotheses. It may be of interest to find out which variables in such a correlational analysis are most important, with some more or less informed expectation about a meaningful ordering in the sizes of the controlled associations, and that some other controlled correlations are equal to each other.

In this paper a general framework is presented for testing multiple hypotheses with equality and order constraints on measures of association. The current approach builds upon the earlier work of \cite{Mulder:2016} who presented a test for order constraints on bivariate correlations between continuous variables. The current paper proposes several important extensions. First, the proposed method can also be used for testing hypotheses with equality constraints, hypotheses with order constraints, as well as hypotheses with combinations of equality and order constraints on the correlations. The importance of this extension is evident given the importance of the (equality constrained) null hypothesis in scientific research, e.g., an association equals zero, or the association between all variables is exactly equal. Table \ref{tablehyp} gives several examples of hypothesis tests that can be executed using the proposed methodology. Note that a Bayesian test for a precise hypothesis was considered by \cite{Wetzels:2012} and a Bayesian test for order-constrained hypotheses was considered by \cite{Mulder:2016}. The table shows that a much broader class of hypothesis tests can be executed using the proposed methodology. Furthermore it will be shown that different priors are needed than the priors that were proposed in these previous papers. Second, the methodology can be used for testing the association between continuous variables, ordinal variables, and combinations of ordinal and continuous variables. This is particularly relevant in sociological research where variables are commonly measured on an ordinal scale. Table \ref{tablecorr} shows an overview of the types of association measures that can be tested using the proposed methodology. Third, the methodology can be used to test constraints on partial correlations by correcting for external covariates. Controlling for external (confounding) covariates is very important to avoid spurious relationships between the variables of interest.

\begin{table}[t]
\caption{Examples of possible tests that can be executed using the proposed methodology.}
{\small \begin{tabular}{ll}
  \hline
 & Example hypothesis test\\
  \hline
Precise testing & $H_0:\rho=0$ versus $H_1:\rho \not =0$\\
One-sided testing & $H_0:\rho\le0$ versus $H_1:\rho \not >0$\\
Multiple hypothesis testing & $H_0:\rho=0$ versus $H_1:\rho <0$ versus $H_2:\rho >0$\\
Interval testing & $H_0:|\rho|\le.1$ versus $H_1:|\rho| >.1$\\
Equality-constrained testing & $H_1:\rho_{12}=\rho_{13}=\rho_{14}$ versus $H_2:\text{``not $H_1$''} $\\
Order-constrained testing & $H_1:\rho_{12}<\rho_{13}<\rho_{14}$ versus $H_2:\rho_{12}>\rho_{13}>\rho_{14}$\\
&versus $H_3:\text{ ``neither $H_1$, nor $H_2$''}$\\
Hypotheses with equality & $H_1:\rho_{12}<\rho_{13}=\rho_{14}$ versus $H_2:\text{ ``not $H_1$''}$\\
and order constraints & \\
\hline
\end{tabular}}
\label{tablehyp}
\end{table}

When testing statistical hypotheses on measures of association, different testing criteria can be used. The Fisherian $p$ value is perhaps most commonly used for testing a single correlation. In this paper however we shall not develop $p$ values for testing multiple hypotheses with equality and/or order constraints on the measures of association. There are two important reasons for this. First, $p$ values cannot be used to directly test nonnested hypotheses with order constraints, such as $H_1:\rho_{21}<\rho_{31}<\rho_{32}$ versus $H_2:\rho_{21}>\rho_{31}>\rho_{32}$; instead $p$ values can only be used for testing a precise null hypothesis with only equality constraints against an ordered alternative or for testing an ordered null hypothesis against the unconstrained alternative \citep{Silvapulle:2004}. Second, $p$ value tests are not suitable for testing multiple hypotheses directly against each other; instead a $p$ value significance test is designed to test one specific null hypothesis (possibly consisting of multiple constraints) against one unconstrained alternative. It would not be possible however to test multiple hypotheses directly against each other, such as $H_0:\rho_{21}=\rho_{31}=\rho_{32}$ versus $H_1:\rho_{21}>\rho_{31}>\rho_{32}$ versus $H_2:\rho_{21}<\rho_{31}<\rho_{32}$ versus the complement hypothesis $H_3:\text{``not $H_0,H_1,H_2$''}$. Furthermore multiple post-hoc tests also have fundamental disadvantages as they may result in conflicting conclusions, e.g., both $\rho_{21}=\rho_{31}$ and $\rho_{21}=\rho_{32}$ are rejected but $\rho_{31}=\rho_{32}$ is not. Another important limitation is that when comparing a $p$ value with a prespecified significance level $\alpha$, it is still possible to reject a true null hypothesis, typically with a probability of $\alpha=.05$, even in the case of extremely large samples as often observed in social research \citep{Raftery:1995}. Due to this inconsistent behavior, classical $p$ values are thus of very limited use.

\begin{table}[t]
\caption{Types of measures of association depending on the measurement scale of the variables.}
{\small \begin{tabular}{lccc}
  \hline
  % after \\: \hline or \cline{col1-col2} \cline{col3-col4} ...
  & \multicolumn{3}{c}{Scale of $Y_2$}\\  
    \cline{2-4}
 & Dichotomous & Polychotomous- & Continuous-\\
Scale $Y_1$& & Ordinal categories & Interval\\
\hline
Dichotomous & Tetrachoric & Polychoric & Biserial\\
\\
Polychotomous- & & Polychoric & Polyserial\\
Ordinal categories & & & \\
\\
Continuous- & & & Product-\\
Interval & & & Moment\\
  \hline
\end{tabular}}
\label{tablecorr}
\end{table}

Another important class for testing statistical hypotheses is the class of information criteria. Well-known examples are the AIC \citep{Akaike:1973}, the BIC \citep[which is based on a large sample approximation of the marginal likelihood, see][]{Schwarz:1978,Raftery:1995}, and the DIC \citep{Spiegelhalter:2002}. These information criteria explicitly balance between fit and complexity when quantifying which model or hypothesis is best for the data at hand. Although information criteria can straightforwardly be used for testing multiple hypotheses or models simultaneously (unlike classical p-values), these criteria are not suitable for testing order constrained hypotheses. The reason is that these information criteria incorporate the complexity of a model based on the number of free parameters which is ill-defined when order constraints are present. As an example, it is unclear how many free parameters are present under the order-constrained hypothesis $H_1:\rho_{21}<\rho_{31}<\rho_{32}$. Model evaluation criteria from the SEM literature, such as the comparative fit index (CFI), are also not suitable for testing order hypotheses for similar reasons \citep{Braeken:2015}. Therefore these model evaluation criteria will not be considered further in this paper.

The criterion that will be used is the Bayes factor \citep{Jeffreys,Kass:1995}. The Bayes factor is the Bayesian quantification of the relative evidence in the data between two competing hypotheses. Thus, the Bayes factor can also be used to quantify evidence for a null hypothesis, which is not the case for the $p$ value; the $p$ value can only be used to falsify the null. Another important property of the Bayes factor is that it can straightforwardly be used for testing multiple hypotheses simultaneously \citep[e.g.,][]{Berger:1999}. In addition Bayes factors can be transformed to so-called posterior probabilities of the hypotheses. These posterior probabilities provide a direct answer to the research question how plausible (in a Bayesian sense) each hypothesis is based on the observed data. Bayes factors are also particularly suitable for testing hypotheses with order constraints. This has been shown by \cite{Klugkist:2005} for group means in ANOVA designs, \cite{Mulder:2009} for repeated measurements, \cite{Mulder:2010} for multivariate regression models, \cite{Klugkist:2010} for contingency tables, \cite{BoeingMessing:2018} for group variances, \cite{MulderFox:2018} for intraclass correlations, and \cite{Gu:2014} for general statistical models. See also \cite{Hoijtink:2011} for an overview of various methods for order-constrained inference using the Bayes factor. Tutorial papers for readers who are new to Bayes factors are, among others, \cite{Hoijtink:2019}, \cite{Rouder:2016}, or \cite{Schoot:2011b}. Finally it is important to note that under very general conditions, Bayes factors and posterior probabilities are consistent which implies that the evidence for the true hypothesis goes to infinity as the sample size grows to infinity. Alternative testing criteria such as the $p$ value or the AIC are not consistent \citep[e.g.][]{OHagan:1995,Berger:1999}.

In order to compute the Bayes factor for testing a set of hypotheses with constraints on the correlations, two challenges need to be overcome. The first challenge is the specification of the prior of the parameters under each hypothesis. The prior plays a central role in a Bayesian analysis and reflects which values of the correlations are most likely a priori. The choice of the prior is particularly important when testing hypotheses with equality constraints on the model parameters \citep{Lindley:1957,Bartlett:1957,Jeffreys}. For this reason arbitrarily specified priors should not be used. In this paper we propose a prior specification method for the correlations under the hypotheses based on uniform distributions. This prior assumes that every combination of the correlations under each hypothesis is equally likely, which seems a reasonable default choice that reflects `prior ignorance'. Note that \cite{Jeffreys} originally proposed a default Bayes factor for testing a single bivariate correlation using a uniform prior \citep{Ly:2016,Ly:2018}. From this point of view the proposed methodology can be seen as a generalization of Jeffreys' original approach. 

The second challenge is the computation of the marginal likelihood, a key ingredient of the Bayes factor. To compute the marginal likelihood of a hypothesis we need to compute the integral of the product of the likelihood and prior over the parameter space of the free parameters. In complex settings the computation of this integral can take a lot of time, which limits general utilization of the methodology for applied users. To tackle this problem we first present a general expression of a Bayes factor for a hypothesis with equality and order constraints on the parameters of interest versus an unconstrained model. This general result is used for the current problem of testing inequality and order constraints on measures of association. Subsequently an accurate approximation of the unconstrained posterior for the measures of association is obtained using an efficient MCMC algorithm that combines several novel techniques on Bayesian computation for the generalized multivariate probit model we consider in this paper. The combination of ordinal and continuous outcome variables is modeled using the model of \cite{Boscardin:2008}. Splitting the covariance matrix in standard deviations and measures of association is achieved by applying the separation strategy of \cite{Barnard:2000}. Posterior correlation matrices are efficiently sampled in one step using the method of \cite{Liu:2006}. To improve mixing of the threshold parameters in the posterior, which can be a serious problem in Bayesian ordinal regression, the parameter expansion of \cite{LiuSabatti:2000} is extended to the generalized multivariate probit model. Finally to simplify the computation of the Bayes factor, the unconstrained posterior is accurately approximated using a multivariate normal distribution after Fisher transformation on the sampled measures of association \citep{Mulder:2016}. The algorithm for computing Bayes factors and the posterior probabilities for the hypotheses based on the new methodology is implemented in a Fortran software program called `{\tt BCT}' (Bayesian Correlation Testing). The program allows users to test a general class of equality and order constrained hypotheses on measures of association which are commonly observed in social research. A user manual is included.

This paper is structured as follows. In the next section we illustrate some equality and order constrained hypotheses that we develop for associations between life and domain satisfaction as addressed in \textit{Quality of Life} research. Then we explain the methodology and numerical computational details of the method, and report evidence of the methods' performance using a small simulation study. We then return to our empirical example and discuss the results of testing the equality and order constrained hypotheses we developed. Finally, we conclude with summarizing the properties and advantages of applying the method to social science research problems.

\section{Empirical Examples}
\subsection{Example 1: Associations between life, leisure and relationship satisfaction}\label{Example1}
\begin{figure}
	\begin{center}
		\includegraphics[width=12cm,keepaspectratio=true]{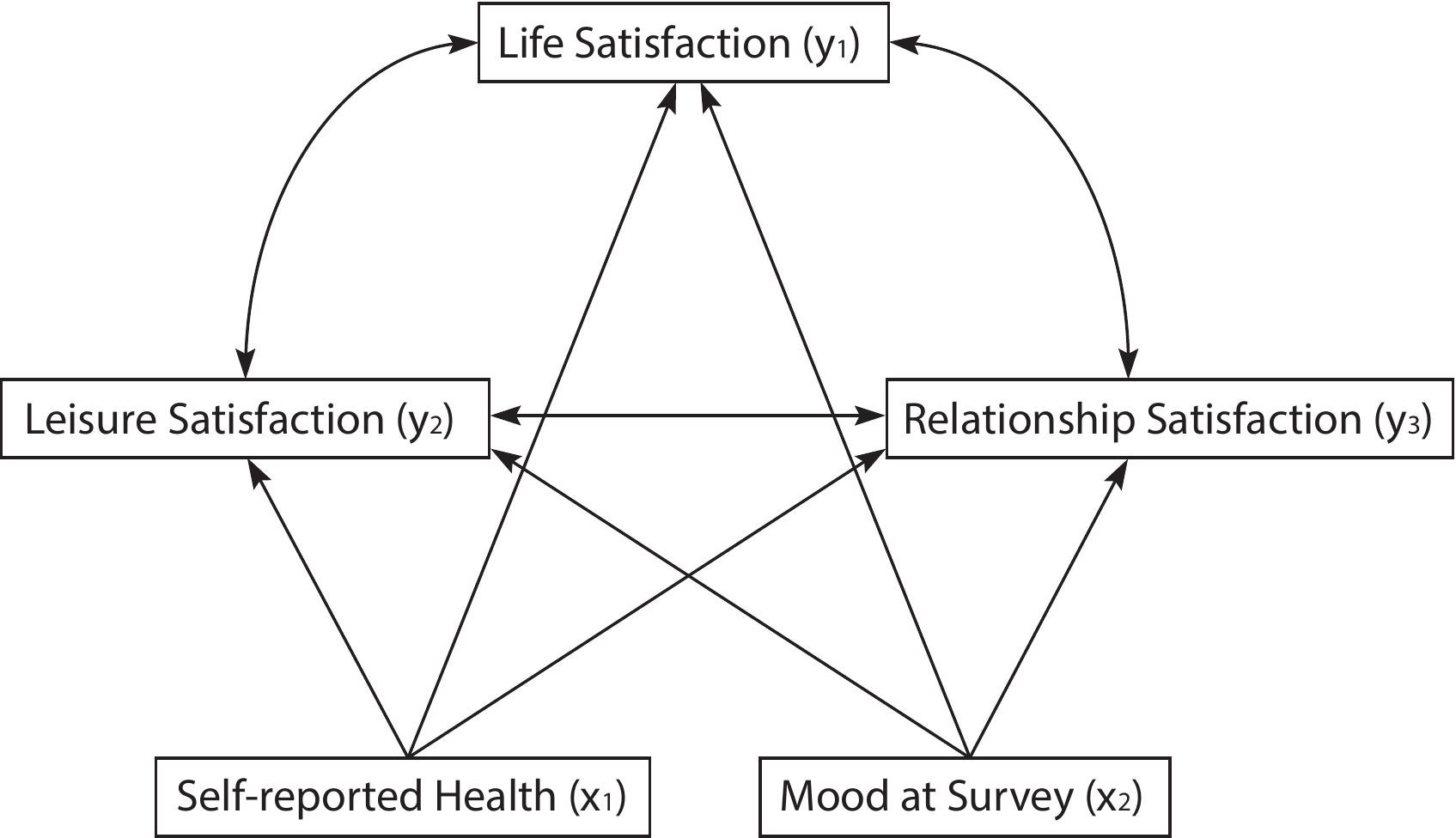}
	\end{center}
\caption{Graphical Model describing Partial Associations between Life and Domain Satisfaction Variables.}
\label{figmodel1}
\end{figure}

An important research question in \textit{Quality of Life} research concerns how a person's satisfaction in certain life domains (e.g.  about leisure, relationship, or work) relates to overall life satisfaction \cite[]{FelcePerry:1995,NewmanTayDiener:2014}. A prevailing hypothesis is that women are much more relational than men \cite[]{Hook:2003}. As these authors point out, women like being connected (i.e., experiencing ``we-ness''), doing things together with others, and they place great emphasis on talking and emotional sharing. Men, on the other hand, see togetherness more as an activity than a state of being, as it is for women. They favor interactions that involve `doing' rather than `being.' Unlike women, men ``prefer to have an element of separation included in their relationships with others and the `doing' orientation seems to promote this'' \cite[p.~465]{Hook:2003}. Research has furthermore shown that men have on average more leisure time than women \cite[]{Mattingly:2003}. Based on these general theoretical ideas and observations concerning gender differences in relational issues and leisure, we anticipate that such systematic differences will also be observed in how satisfaction with one's relationship and leisure is associated with overall life satisfaction, with leisure satisfaction relating more strongly to overall life satisfaction among men and relationship satisfaction relating more strongly to overall life satisfaction among women. Consider the following graphical model about the partial associations between three focal dependent variables that might be included in such an analysis: the degree of life satisfaction, leisure satisfaction, and relationship satisfaction (Figure \ref{figmodel1}). In this example we are interested in testing various informative hypotheses about conditional partial associations between the variables concerned, with gender also potentially moderating such partial associations. As can be seen, in this model the associations between the key variables of interest are controlled for differences in self-reported health and mood at survey.

We use data from the Dutch LISS panel (Longitudinal Internet Studies for the Social sciences) to test various informative hypotheses about this model. In particular, the LISS panel contains the following variables used to operationalize the variables in the conceptual model:

\begin{enumerate}
	\item  \textit{Life Satisfaction} ($y_1$): respondents were asked to rate the following statement: ``I am satisfied with my life''; ordered categorical variable, with 1 = strongly disagree -- 7 strongly agree
	
	\item  \textit{Leisure Satisfaction} ($y_2$): respondents were asked to indicate how satisfied they are with the way in which they spend their leisure time; continuous variable, with 1 = entirely dissatisfied -- 11 entirely satisfied
	
	\item  \textit{Relationship Satisfaction} ($y_3$): respondents were asked to indicate how satisfied they are with their current relationship; continuous variable, with 1 = entirely dissatisfied -- 11 entirely satisfied
	
	\item  \textit{Self-reported Health} ($x_1$):  respondents were asked: ``How would you describe your health, generally speaking?''; ordered categorical variable, with 1 = poor -- 5 = excellent
	
	\item  \textit{Mood at Survey }($x_2$): respondents were asked to indicate how they feel at the moment of completing the survey; ordered categorical variable, with 1 = very bad and 7 = very good
	
	\item  \textit{Gender }(grouping variable $g$): the respondent's gender, with 1 = men and 2 = women
\end{enumerate}

\noindent Several competing informative hypotheses can be formulated about the ordering of, and equalities between the partial correlations  between \textit{y1}, \textit{y2} and \textit{y3} conditional on \textit{g}. Here, we consider the following hypotheses:\\
\\
\noindent\textit{Hypotheses about partial associations between satisfaction domains for men:}

\begin{description}
	\item[Hypothesis $H_{1_a}:$] The partial association between leisure satisfaction and life satisfaction is stronger than the partial association between relationship satisfaction and life satisfaction, and in turn this association is stronger than the partial association between leisure satisfaction and relationship satisfaction: ${\mathrm{\rho }}_{g1y2y1}>{\mathrm{\rho}}_{g1y3y1}>{\mathrm{\rho}}_{g1y3y2}$.
	
	\item[Hypothesis $H_{1_b}:$] The partial association between relationship satisfaction and life satisfaction is stronger than the partial association between leisure satisfaction and life satisfaction, and in turn this association is stronger than the partial association between leisure satisfaction and relationship satisfaction: ${\mathrm{\rho }}_{g1y3y1}>{\mathrm{\rho }}_{g1y2y1}>{\mathrm{\rho }}_{g1y3y2}$.

\item[Hypothesis $H_{1_c}:$] The partial associations between life satisfaction, leisure satisfaction, and relationship satisfaction are equal:  ${\mathrm{\rho }}_{g1y3y1}={\mathrm{\rho }}_{g1y2y1}={\mathrm{\rho }}_{g1y3y2}$.

\item[Hypothesis $H_{1_d}:$] The complement hypothesis, which implies `not $H_{1_a},H_{1_b},H_{1_c}$'.
\end{description}

\noindent \textit{Similar hypotheses about partial associations between satisfaction domains for women:}

\begin{description}
	\item[Hypothesis $H_{2_a}:$] ${\mathrm{\rho }}_{g2y2y1}>{\mathrm{\rho}}_{g2y3y1}>{\mathrm{\rho}}_{g2y3y2}$.
	
	\item[Hypothesis $H_{2_b}:$] ${\mathrm{\rho }}_{g2y3y1}>{\mathrm{\rho }}_{g2y2y1}>{\mathrm{\rho }}_{g2y3y2}$.

\item[Hypothesis $H_{2_c}:$] ${\mathrm{\rho }}_{g2y3y1}={\mathrm{\rho }}_{g2y2y1}={\mathrm{\rho }}_{g2y3y2}$.

\item[Hypothesis $H_{2_d}:$] The complement hypothesis, which implies `not $H_{2_a},H_{2_b},H_{2_c}$'.
\end{description}

\noindent \textit{Hypotheses about conditional partial associations between life satisfaction and leisure satisfaction, with gender as moderator:}

\begin{description}
\item [Hypothesis $H_{3_a}:$] The partial association between leisure satisfaction and life satisfaction among men is stronger than the partial association among women: ${\mathrm{\rho }}_{g\mathrm{1}y\mathrm{2}y\mathrm{1}}\mathrm{>}{\mathrm{\rho }}_{g\mathrm{2}y\mathrm{2}y\mathrm{1}}$.

\item [Hypothesis $H_{3_b}:$] The partial association between leisure satisfaction and life satisfaction among men is equally strong as the partial association among women: ${\mathrm{\rho }}_{g\mathrm{1}y\mathrm{2}y\mathrm{1}}\mathrm{=}{\mathrm{\rho }}_{g\mathrm{2}y\mathrm{2}y\mathrm{1}}$.

\item [Hypothesis $H_{3_c}:$] The partial association between leisure satisfaction and life satisfaction among women is stronger than the partial association among men: ${\mathrm{\rho }}_{g\mathrm{1}y\mathrm{2}y\mathrm{1}}\mathrm{<}{\mathrm{\rho }}_{g\mathrm{2}y\mathrm{2}y\mathrm{1}}$.

\end{description}

\noindent \textit{Hypotheses about conditional partial associations between relationship satisfaction and life satisfaction, with gender as moderator:}

\begin{description}

\item [Hypothesis $H_{4_a}:$] The partial association between relationship satisfaction and life satisfaction among women is stronger than the partial association among men: ${\mathrm{\rho }}_{g\mathrm{2}y\mathrm{3}y\mathrm{1}}\mathrm{>}{\mathrm{\rho }}_{g\mathrm{1}y\mathrm{3}y\mathrm{1}}$.

\item [Hypothesis $H_{4_b}:$] The partial association between relationship satisfaction and life satisfaction among men is equally strong as the partial association among women: ${\mathrm{\rho }}_{g\mathrm{2}y\mathrm{3}y\mathrm{1}}\mathrm{=}{\mathrm{\rho }}_{g\mathrm{1}y\mathrm{3}y\mathrm{1}}$.

\item [Hypothesis $H_{4_c}:$] The partial association between relationship satisfaction and life satisfaction among men is stronger than the partial association among women: ${\mathrm{\rho }}_{g\mathrm{2}y\mathrm{3}y\mathrm{1}}\mathrm{<}{\mathrm{\rho }}_{g\mathrm{1}y\mathrm{3}y\mathrm{1}}$.

\end{description}

\subsection{Example 2: Association between egalitarian justice beliefs across countries }
Two important egalitarian norms that people can use in evaluating how societal resources (e.g., wealth, public goods) should be distributed are the principle of equality and the principle of need \cite{deutsch1975equity}. When people apply the equality principle, they support the idea that everyone should receive the same amount of such valuable societal resources. When people use the principle of need, they believe that societal resources should be distributed according to an individual’s needs, for example the minimum resources required to lead a decent life. 

As \citet[p. 424]{smith2012two} point out, the principles of need and equality are similar in their egalitarian belief and make that both justice principles are positively correlated (Arts and Gelissen, 2001), the degree to which these justice principles are positively associated presumably varies between countries. In this respect, in Europe, a critical socio-historical divide exists between the Western European countries and the post-communist countries of Eastern Europe. In the context of communism, egalitarianism was the dominant ideology, and we can expect that citizens of former communist countries are only slowly abandoning strong endorsement of egalitarian beliefs even in the light of market reforms \citep[p. 425]{smith2012two}. Therefore, we expect that the correlation between both egalitarian justice principles will be stronger in post-communist countries than in Western European countries; this will hold even if in Western European countries public endorsement of the egalitarian ideology also has played an important role but within the context of a long-established capitalist market economy. Among the publics of capitalist Western European countries, presumably, equity or meritocratic considerations (i.e., assigning societal resources proportional to each’s contributions or merits that deserve reward) also play a crucial role in deciding how societal resources should be distributed. Consequently, endorsement of egalitarian justice principles will be more fragmented in Western European countries which presumably results in weaker associations between the need and equality principle. Of course, even in Western European countries there is considerable difference in the degree to which egalitarian principles are institutionally embedded, with the Swedish welfare state being the model of the social-democratic welfare state. Finally, we expect that such differences in the strength of (partial) association between both egalitarian principles hold across countries, even if we control for variables such as gender, age and educational attainment, which relate to such justice principles \citep{arts2001welfare}.

We use data from the European Values Survey 2000 to test informative hypotheses about the presumed cross-national differences in the association between support for the equality and need principle. We limit our investigation to four countries that reflect the difference between post-communist countries and Western European capitalist countries with a more or less egalitarian ideology: Bulgaria, Romania, The Netherlands, and Sweden. We use the following variables:

\begin{enumerate}

\item	``In order to be considered ‘just’, what should a society provide? Please tell me for each statement if it is important or unimportant to you.''

\textit{Endorsement of the equality principle} ($y_1$): ``Eliminating big inequalities in income between citizens''; ordered categorical variable, with 1 = not at all important – 5 very important

\textit{Endorsement of the need principle} ($y_2$): ``Guaranteeing that basic needs are met for all, in terms of food, housing, clothes, education, health''; ordered categorical variable, with 1 = not at all important – 5 very important
\item	\textit{Gender} ($x_1$):  the respondent’s gender, with 1 = men and 2 = women, 
\item	\textit{Age} ($x_2$): the respondent’s age in years
\item 	\textit{Education} ($x_3$): age at which full-time education will or was completed (continuous variable with eleven intervals, ranging from 0 = no formal education to 10 = 21 and more years)
\item	\textit{Country} (grouping variable $g$): the respondent’s country, with 1= Bulgaria, 2 = Romania, 3 = Sweden, and 4 = The Netherlands
\end{enumerate}

We formulate the following informative hypotheses:

\begin{description}

\item[Hypothesis $H_{5_a}:$] The partial association between support for the equality principle and the need principle is ordered as follows: it is strongest in Bulgaria, weaker in Romania (because of more violent resistance against communist rule and dictatorship in this country), even weaker in Sweden (Western European country with a strong egalitarian welfare ideology) and weakest in The Netherlands (a Western European country with a moderately egalitarian welfare ideology) : \\ ${\rho _{g1y2y1}} > {\rho _{g2y2y1}} > {\rho _{g3y2y1}} > {\rho _{g4y2y1}}$

\item[Hypothesis $H_{5_b}:$] The partial association between support for the equality principle and the need principle is equal between all countries: \\
${\rho _{g1y2y1}} = {\rho _{g2y2y1}} = {\rho _{g3y2y1}} = {\rho _{g4y2y1}}$

\item[Hypothesis $H_{5_c}:$] The partial association between support for the equality principle and the need principle is equal among post-communist countries Bulgaria and Romania, and the partial association between support for the equality principle and the need principle is equal among the Western European countries Netherlands and Sweden; however, the association between support for the equality principle and the need principle is stronger among post-communist countries than among Western European countries:
${\rho _{g1y2y1}} = {\rho _{g2y2y1}} > {\rho _{g3y2y1}} = {\rho _{g4y2y1}}$

\item[Hypothesis $H_{5_d}:$] The complement hypothesis, which implies `not $H_{5_a}$, $H_{5_b}$, $H_{5_c}$'.

\end{description}

\section{Model specification}
\subsection{The generalized multivariate probit model}\label{sectionmodel}
A generalized multivariate probit regression model will be used for modeling combinations of continuous variables and ordinal dependent variables. This model is well-established in the Bayesian literature \citep[e.g.,][]{AlbertChib:1995, ChibGreenberg:1998,ChenDey:2000,LiuSabatti:2000,Barnard:2000,Kotas:2005,Fox:2005,Raach:2005,Fahrmeir:2007,Lawrence:2008,Boscardin:2008,Asparouhov:2010}. We consider $G$ independent populations. In the first empirical application discussed above for example there were $G=2$ populations: a male population and a female population. The $P$-dimension vector of dependent variables of subject $i$ in population $g$ will be denoted by $\textbf{y}_{ig}'=(\textbf{v}_{ig}',\textbf{u}'_{ig})$, of which the first $P_1$ elements, $\textbf{v}_{ig}'=(v_{ig1},\ldots,v_{igP_1})$, are continuous normally distributed variables and the remaining $P_2=P-P_1$ elements, $\textbf{u}_{ig}'=(u_{ig1},\ldots,u_{igP_2})$, are measured on an ordinal scale, for $i=1,\ldots,n_g$, and $g=1,\ldots,G$. Furthermore, we assume that the $p$-th ordinal variable can assume the categories $1,\ldots,K_p$, for $p=1,\ldots,P_2$.

As is common in a Bayesian multivariate probit modeling, a multivariate normal latent variable, denoted by $\textbf{z}_{ig}$, is used for each ordinal vector $\textbf{u}_{ig}$. This implies that the $p$ ordinal variable of subject $i$ in population $g$ falls in category $k$, i.e.,
\[
u_{igp}=k,~\mbox{if}~z_{igp}\in(\gamma_{gp(k-1)},\gamma_{gpk}],
\]
for $k=1,\ldots,K_p$, where $\gamma_{gpk}$ is the upper cut-point of the $k$-th category of the $p$-th ordinal variable in the $g$-th population. To ensure identification of the model it is necessary to set $\gamma_{gp0}=-\infty$, $\gamma_{gp1}=0$, and $\gamma_{gpK_p}=\infty$, for $g=1,\ldots,G$ and $p=1,\ldots,P_2$, and to fix the error variances of the latent variables to 1 as is common in multivariate probit modeling \citep{ChibGreenberg:1998}.

The mean structure of each dependent variable is assumed to be a linear combination of $Q$ external covariates $\textbf{x}_{ig}$. Subsequently, the generalized multivariate probit model can be defined by
\begin{eqnarray}
\label{model1}\left[\begin{array}{c}
\textbf{v}_{ig}\\
\textbf{z}_{ig}
\end{array}\right]
&\sim& N(\textbf{B}_g\textbf{x}_{ig},\bm\Sigma_g),~\mbox{where}\\
\label{model2}\bm\Sigma_g&=&\mbox{diag}(\bm\sigma_g',\textbf{1}'_{P_2})\textbf{C}_g\mbox{diag}(\bm\sigma_g',\textbf{1}'_{P_2}),
\end{eqnarray}
where $\textbf{1}'_{P_2}$ is a vector of length $P_2$ of ones. Notice here that the correlation matrices $\textbf{C}_g$ and standard deviations $\bm\sigma_g$ are separately modeled as in \cite{Barnard:2000}. In this model $\textbf{B}_g$ is a $P\times Q$ matrix with regression coefficients of the $g$-th population where element $(p,q)$ reflects the effect of the $q$-th covariate on the $p$-th dependent variable, for $q=1,\ldots,Q$ and $p=1,\ldots,P$, the $P_1$ error standard deviations of $\textbf{v}_{ig}$ in population $g$ are contained in $\bm\sigma_g$, and $\textbf{C}_g$ denotes the $P\times P$ correlation matrix of population $g$. Note that the $(p_1,p_2)$-th element of $\textbf{C}_g$ denotes the linear association of the $p_1$-th and $p_2$-th dependent variable in population $g$ while controlling for the covariates in $\textbf{x}_{ig}$. Thus, if a model is specified with two dependent variables and several covariates, and we would be interested in testing $\rho_{12}=0$, we are essentially testing the partial correlation between the two dependent variables while controlling for the covariates that are included in $\textbf{x}_{ig}$. This way no distributional assumptions are made about the covariates because they are included as independent variables in the regression model. For example one can include covariates that are only 0 or 1 to correct for a categorical variable. If $Q=1$ and $x_{ig}=1$, no covariates are incorporated in the model, which implies that the elements in $\textbf{C}_{g}$ are bivariate correlations.

\subsection{Hypothesis testing on measures of association}
In the context of testing constraints on correlations, which is the goal of the current paper, the correlation matrices $\textbf{C}_g$ are of central importance while the parameter matrix $\textbf{B}_g$ and the variances $\bm\sigma$ are treated as nuisance parameters. The correlations in $\textbf{C}_g$ are contained in the vector $\bm\rho$, e.g.,
\begin{eqnarray}
\textbf{C}_g = \left[\begin{array}{ccc}
1\\
\rho_{g21} & 1\\
\rho_{g31} & \rho_{g32} & 1
\end{array}\right] \Rightarrow \bm\rho_g=(\rho_{g21},\rho_{g31},\rho_{g32})'.
\label{corr3}
\end{eqnarray}
Furthermore, all correlations in the $G$ different correlation matrices, $\textbf{C}_g$, will be combined in the vector $\bm\rho'=(\bm\rho_1',\ldots,\bm\rho_G')$ of length $L=\frac{1}{2}GP(P-1)$. Similarly we combine the parameter matrices $\textbf{B}_g$ and variances $\bm\sigma_g$ over all $G$ population in the matrices $\textbf{B}$ and $\bm\sigma$, respectively, and subsequently, the vectorization of $\textbf{B}$ is denoted by the vector $\bm\beta$.

We consider a multiple hypothesis test of $T$ hypotheses $H_1,\ldots,H_T$ of the form
\begin{equation}
H_t:\textbf{R}_t^E\bm\rho=\textbf{r}_t^E,~\textbf{R}_t^I\bm\rho>\textbf{r}_t^I,
\label{Ht}
\end{equation}
for $t=1,\ldots,T$, where $[\textbf{R}_t^E|\textbf{r}_t^E]$ is a matrix of coefficients that capture the set of equality constraints under $H_t$ and $[\textbf{R}_t^I|\textbf{r}_t^I]$ is a matrix of coefficients that capture the set of inequality (or order) constraints under $H_t$. In most applications researchers either compare two correlations with each other, e.g., $\rho_{121}>\rho_{131}$, or a single correlation is compared to constant, e.g., $\rho_{121}>.5$. Therefore each row of the matrices $[\textbf{R}_t^E]$ and $[\textbf{R}_t^I]$ is either a permutation of $(1,-1,0\ldots,0)$ with corresponding constant in $\textbf{r}_t^E$ and $\textbf{r}_t^I$ equals 0, or a row is a permutation of $(\pm 1,0,\ldots,0)$ with corresponding constant $r\in(-1,1)$. As an example, hypothesis $H_{1_a}:\rho_{121}>\rho_{131}>\rho_{132}$ from Section \ref{Example1} (with the index labels omitted) would have the following matrix form
\[
\textbf{\textit{H1}$_a$}: \left[
\begin{array}{cccccc}
1 & -1 & 0 & 0 & 0 & 0\\
0 & 1 & -1 & 0 & 0 & 0
\end{array}
\right]\left[
\begin{array}{c}
\rho_{121}\\ \rho_{131}\\ \rho_{132}\\ \rho_{221}\\ \rho_{231}\\ \rho_{232}
\end{array}\right] > \left[
\begin{array}{c}
0 \\ 0
\end{array}\right].
\]

Throughout the paper the allowed parameter space under the constrained $H_t$ will be denoted by $\mathcal{C}_t$. In certain parts of the paper we refer to an unconstrained hypothesis, denoted by $H_u$, which does not assume any constraints on the correlations besides the necessary constraints on $\bm\rho$ that ensure that the corresponding correlation matrices are positive definite.

\section{Methodology}
\subsection{Marginal likelihoods, Bayes factors, and posterior probabilities}
The Bayes factor is a Bayesian criterion that quantifies the relative evidence in the data between two hypotheses. The Bayes factor of hypothesis $H_1$ versus $H_2$ is defined by the ratio of the marginal likelihoods under the respective hypotheses, i.e.,
\begin{equation}
\label{BF1}
B_{12}=\frac{m_1(\textbf{Y})}{m_2(\textbf{Y})},
\end{equation}
where the marginal likelihood is defined by
\begin{equation}
\label{marglike}
m_t(\textbf{Y})=\iiint_{\mathcal{C}_t}p_t(\textbf{Y}|\textbf{X},\bm\beta,\bm\sigma,\bm\rho)\pi_t(\bm\beta,\bm\sigma,\bm\rho)d\bm\rho d\bm\sigma d\bm\beta,
\end{equation}
where $p_t$ denotes the likelihood of the data $\textbf{Y}$ given the unknown parameters and the covariates $\textbf{X}$ under $H_t$, which follows directly from \eqref{model1} and \eqref{model2}, and $\pi_t$ denotes the prior for the unknown model parameters under $H_t$. The marginal likelihood captures how likely the observed data is under a hypothesis and its respective prior. %It has been advocated to also use a hypothesis index for the parameters \citep[e.g.][]{Berger1996} because a parameter can have a different meaning under different hypothesis. For example, a negative correlation under, say, $H_1:\rho_{1,21}<0$, clearly has a different meaning than a positive correlation under, say, $H_2:\rho_{1,21}>0$, which could be captured using an additional hypothesis index 1 and 2 for $\rho_{1,21}$ under $H_1$ and $H_2$, respectively. Throughout this paper we shall this additional index to simplify the notation. Also note that the parameters in the marginal likelihood \eqref{marglike} are dummy variables anyway because they are integrated out.

One of the main strengths of the Bayes factor is its intuitive interpretation. For example, the Bayes factor is symmetrical in the sense that if, say $B_{12}=.1$, which implies that $H_1$ receives 10 times less evidence than $H_2$, it follows naturally from \eqref{BF1} that $B_{21}=B_{12}^{-1}=10$, which implies that $H_2$ receives 10 times more evidence from the data than $H_1$. Furthermore, Bayes factors are transitive in the sense that if $H_1$ received 10 times more evidence than $H_2$, i.e., $B_{12}=10$, and $H_2$ received 5 times more evidence than $H_3$, i.e., $B_{32}=5$, it again follow naturally from the definition in \eqref{BF1} that hypothesis $H_1$ received 50 times more evidence than $H_3$ because $B_{13}=B_{12}\times B_{23}=10\times 5=50$.

Various researchers have provided an indication how to interpret Bayes factors \citep[e.g.][]{Jeffreys,Raftery:1995}. For completeness we provided the guidelines that were given by \cite{Raftery:1995}. These guidelines are helpful for researchers who are new to Bayes factors. We do not recommend to use these guidelines as strict rules because researchers should decide by himself or herself when he or she feels that the Bayes factor indicates strong evidence.

\begin{table}[t!]
\centering
\caption{Rough guidelines for interpreting Bayes factors \citep{Raftery:1995}.}
\begin{tabular}{ll}
\hline
$B_{12}$ & Evidence\\
\hline
$< 1/150$   & Very strong evidence for $H_2$\\
1/150 to 1/20 & Strong evidence for $H_2$\\
1/20  to 1/3  & Positive evidence for $H_2$\\
1/3  to 1   & Weak evidence for $H_2$\\
1 & No preference\\
1  to 3   & Weak evidence for $H_1$\\
3  to 20  & Positive evidence for $H_1$\\
20 to 150 & Strong evidence for $H_1$\\
$> 150$   & Very strong evidence for $H_1$\\
\hline
\end{tabular}
\label{2:tab:interpretationbayesfactor}\end{table}

The Bayes factor can be used to update the prior odds of any pair two hypotheses that can be true before observing the data to obtain the posterior odds that the hypotheses are true after observing the data according to
\begin{equation}
\label{postodds}
\frac{\mbox{Pr}(H_1|\textbf{Y})}{\mbox{Pr}(H_2|\textbf{Y})} = B_{12}\times \frac{\mbox{Pr}(H_1)}{\mbox{Pr}(H_2)},
\end{equation}
where $\mbox{Pr}(H_t)$ and $\mbox{Pr}(H_2|\textbf{Y})$ denote the prior and posterior probability that $H_t$ is true, respectively, for $t=1$ or 2. In the general case of $T$ hypotheses, the posterior hypothesis probabilities can be obtained as follows
\begin{equation}
\mbox{Pr}(H_t|\textbf{Y}) = \frac{\mbox{Pr}(H_t)B_{t1}}{\sum_{t'=1}^T\mbox{Pr}(H_{t'})B_{t'1}}.
\label{posthypo}
\end{equation}
Posterior hypothesis probabilities are useful because they provide a direct answer to the research question how plausible each hypothesis is in light of the observed data. Researchers typically find these posterior probabilities easier to interpret than Bayes factors because the posterior probabilities add up to one. It should be noted however that in the case of equal prior probabilities for the hypotheses, i.e., $\mbox{Pr}(H_1)=\mbox{Pr}(H_2)$, which is the default setting, the posterior odds between two hypotheses corresponds exactly to the respective Bayes factor as can be seen in \eqref{postodds}.

\subsection{Prior specification}\label{sectionprior}
In order to compute the marginal likelihoods (and subsequently, the Bayes factors and posterior hypothesis probabilities), prior distributions need to be formulated for the unknown parameters $\bm\beta$, $\bm\sigma$, and $\bm\rho$ under each constrained hypothesis. A prior distribution, or simply prior, reflects the plausibility of the possible values of the free parameters before observing the data.

Under each hypothesis, we set independent priors for the three different types of model parameters, i.e.,
\begin{equation}
\pi_t(\bm\beta,\bm\sigma,\bm\rho)=\pi_t^N(\bm\beta)\times\pi_t^N(\bm\sigma)\times \pi_t^U(\bm\rho)\times I(\bm\rho\in\mathcal{C}_t),
\end{equation}
with noninformative improper priors for the nuisance parameters
\begin{eqnarray}
\label{betaprior}\pi_t^N(\bm\beta)&\propto&1\\
\label{sigmaprior}\pi_t^N(\bm\sigma)&\propto&\prod_{g=1}^G\sigma_{g,1}^{-1}\times\ldots\times\sigma_{g,P_1}^{-1}.%\prod_{j=1,p=1}^{J,P_1}\sigma_{j,p}^{-1}.
\end{eqnarray}
The domains for $\bm\beta$ for $\bm\sigma$ are $\mathbb{R}^{PQ}$ and $\left(\mathbb{R}^+\right)^{P}$, respectively. Note that the priors for the nuisance parameters are equivalent to the commonly used independence Jeffreys priors which are commonly used for a default Bayesian analysis.
These noninformative improper priors are allowed for these common nuisance parameters as the Bayes factor will be virtually independent to the exact choice of these priors as long as the priors are vaguely enough. This will be explained later in this paper.

It is well known that proper priors (i.e., prior distributions that integrate to one) need to be formulated for the parameters that are tested, i.e., the correlations, in order for the Bayes factor to be well-defined. In this paper we consider a uniform prior for the correlations under a hypothesis in its allowed constrained subspace. This implies that every combination of values for the correlations that satisfies the constraints is equally likely a priori under each hypothesis. The prior for $\bm\rho$ is zero outside the constrained subspace under each hypothesis. Because the constrained parameter space $\mathcal{C}_t$ is bounded, a proper uniform prior can be formulated for the correlations under every hypothesis, i.e.,
\begin{equation}
\pi_t^U(\bm\rho) = V_t^{-1}\times I(\bm\rho\in\mathcal{C}_t),
\label{corrprior}
\end{equation}
where the normalizing constant $V_t$ is given by
\begin{equation}
\label{volume}
V_t = \int_{\mathcal{C}_t} 1 d\bm\rho.
\end{equation}
The normalizing constant $V_t$ can be seen as a measure of the size or volume of the constrained space.

To illustrate the priors under different constrained hypotheses on measures, let us consider a model with 3 dependent variables and one population (the correlation matrix was given in \eqref{corr3} where we omit the population index $j$). Furthermore, let us consider the following hypotheses:
\begin{itemize}
\item $H_u:\rho_{21},\rho_{31},\rho_{32}$ (the unconstrained hypothesis).
\item $H_1:\rho_{21}=\rho_{31}=\rho_{32}$ (all $\rho$'s are equal).
\item $H_2:\rho_{31}=0,\rho_{21},\rho_{32}$ (only $\rho_{31}$ is restricted to zero).
\item $H_3:\rho_{31}=0,\rho_{21}>\rho_{32}$ ($\rho_{31}$ is restricted to zero and $\rho_{21}$ is larger than $\rho_{32}$).
\end{itemize}

\begin{figure}[t!]
\centering\includegraphics[width=1\textwidth]{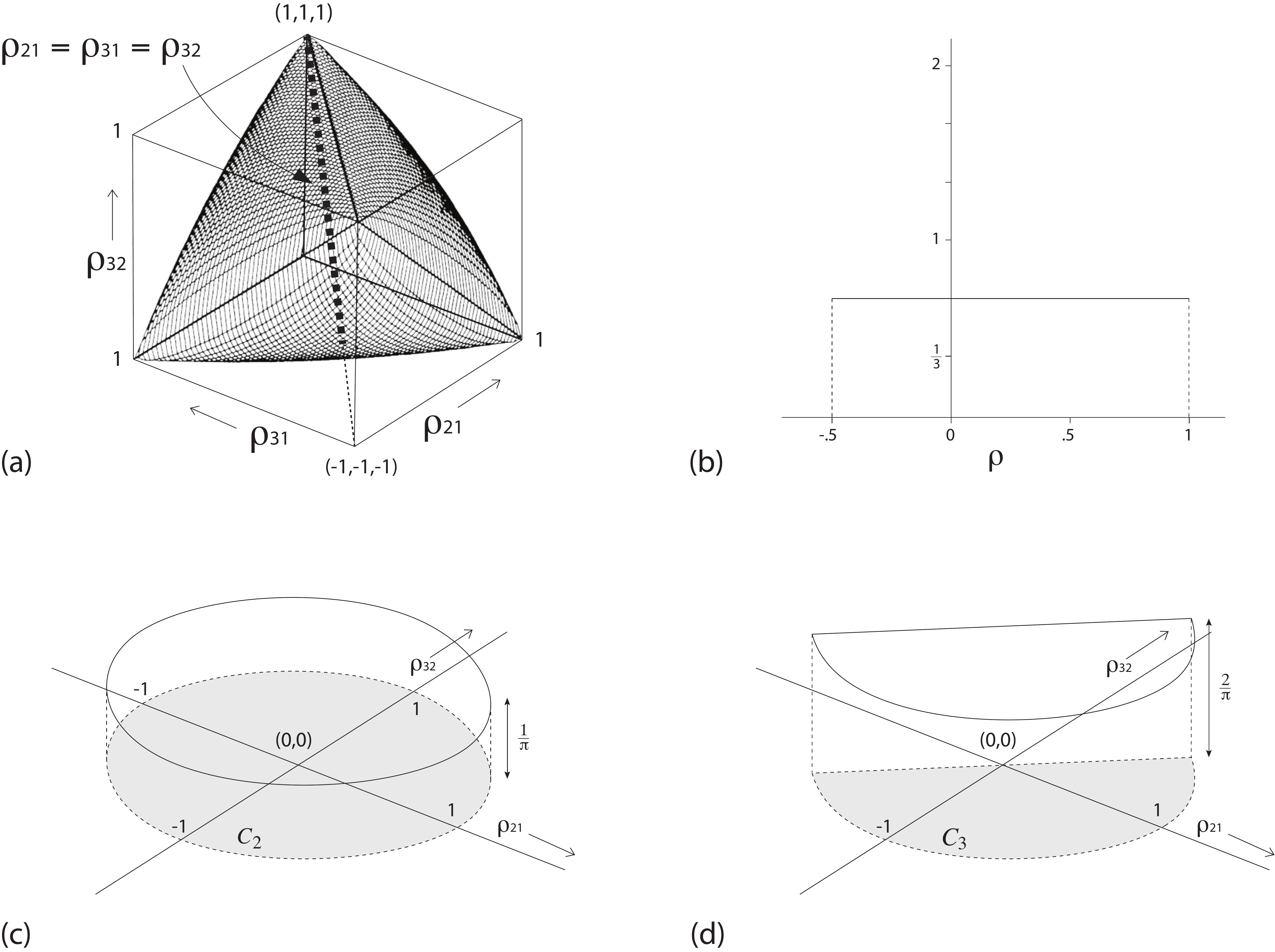}
\caption{(a) Graphical representation of the subspace of $(\rho_{21},\rho_{31},\rho_{32})$ for which the 3-dimensional correlation matrix is positive definite (taken from \cite{Rousseeuw:1994} with permission). The thick diagonal line from $(-\frac{1}{2},-\frac{1}{2},-\frac{1}{2})$ to $(1,1,1)$ represents the correlations that satisfy $\rho_{21}=\rho_{31}=\rho_{32}$ and result in a positive diagonal correlation matrix. (b) Uniform prior $\pi_1^U$ for the common correlation $\rho$ under $H_1:\rho_{21}=\rho_{31}=\rho_{32}$ in the allowed region $\mathcal{C}_1=\{\rho|\rho\in(-\frac{1}{2},1)\}$. (c) Uniform prior $\pi_2^U$ for the free parameters under $H_2:\rho_{31}=0$ in the allowed region $\mathcal{C}_2=\{(\rho_{21},\rho_{32})|\rho_{21}^2+\rho_{32}^2<1\}$. (d) Uniform prior for the free correlations under $H_3:\rho_{31}=0,\rho_{21}>\rho_{32}$ in the allowed region $\mathcal{C}_3=\{(\rho_{21},\rho_{32})|\rho_{21}^2+\rho_{32}^2<1,\rho_{21}>\rho_{32}\}$.}
\label{fig1}
\end{figure}

The unconstrained parameter space of $\bm\rho=(\rho_{21},\rho_{31},\rho_{32})'$ that results in a positive definite correlation matrix is displayed in Figure \ref{fig1}a (taken from \cite{Rousseeuw:1994} with permission). As noted by \cite{Joe:2006}, the volume of this 3-dimensional subspace equals $4.934802$. Therefore, the volume under the unconstrained hypothesis $H_u$ is given by $V_u=4.934802$ in \eqref{volume}. Therefore, the unconstrained uniform prior for the correlations equals $\pi_u^U(\bm\rho) = \frac{1}{4.934802}\times I(\bm\rho\in\mathcal{C}_u)$ in \eqref{corrprior}.

When all $\rho$'s are equal as under $H_1$, the common correlation, say, $\rho$, must lie in the interval $(-\frac{1}{2},1)$ to ensure positive definiteness \citep[e.g.][]{MulderFox:2013}. Thus, the size of the parameter space of $\rho$ corresponds to the length of the interval which is $V_1=\frac{3}{2}$. Therefore, the uniform prior for $\rho$ under $H_1$ corresponds to $\pi_1^U(\rho)=\frac{2}{3}\times I(\rho\in(-\frac{1}{2},1))$, which is plotted in Figure \ref{fig1}b. The diagonal $\rho_{21}=\rho_{31}=\rho_{32}$ is also plotted in Figure \ref{fig1}a as a dashed line, where the thick part lies within positive definite subspace $\mathcal{C}_u$.

When $\rho_{31}$ is restricted to zero as under $H_2$, the allowed parameter space for $(\rho_{21},\rho_{32})$ that results in a positive definite correlation matrix must satisfy $\rho_{21}^2+\rho_{32}^2< 1$, i.e., a circle with radius 1 \citep{Rousseeuw:1994}. Therefore, the uniform prior under $H_2:\rho_{31}=0$ is given by $\pi_2(\rho_{21},\rho_{32})=\frac{1}{\pi}\times I(\rho_{21}^2+\rho_{32}^2<1)$ (Figure \ref{fig1}c) because a circle or radius 1 has a surface of $V_2=\pi$.

When $\rho_{31}$ is restricted to zero and $\rho_{21}$ is larger than $\rho_{32}$ as under $H_3$, the subspace is half as small as under $H_2$. Therefore the uniform prior density for the free parameters under $H_3$ is twice as large as the prior under $H_1$ to ensure the prior integrates to one. Thus, the uniform prior under $H_3$ is given by $\pi_3(\rho_{21},\rho_{32})=\frac{2}{\pi}\times I(\rho_{21}^2+\rho_{32}^2<1~\&~\rho_{21}>\rho_{32})$ (Figure \ref{fig1}d).

\subsubsection{Comparison with other prior choices}
\begin{figure}[t!]
\centering\includegraphics[width=.7\textwidth]{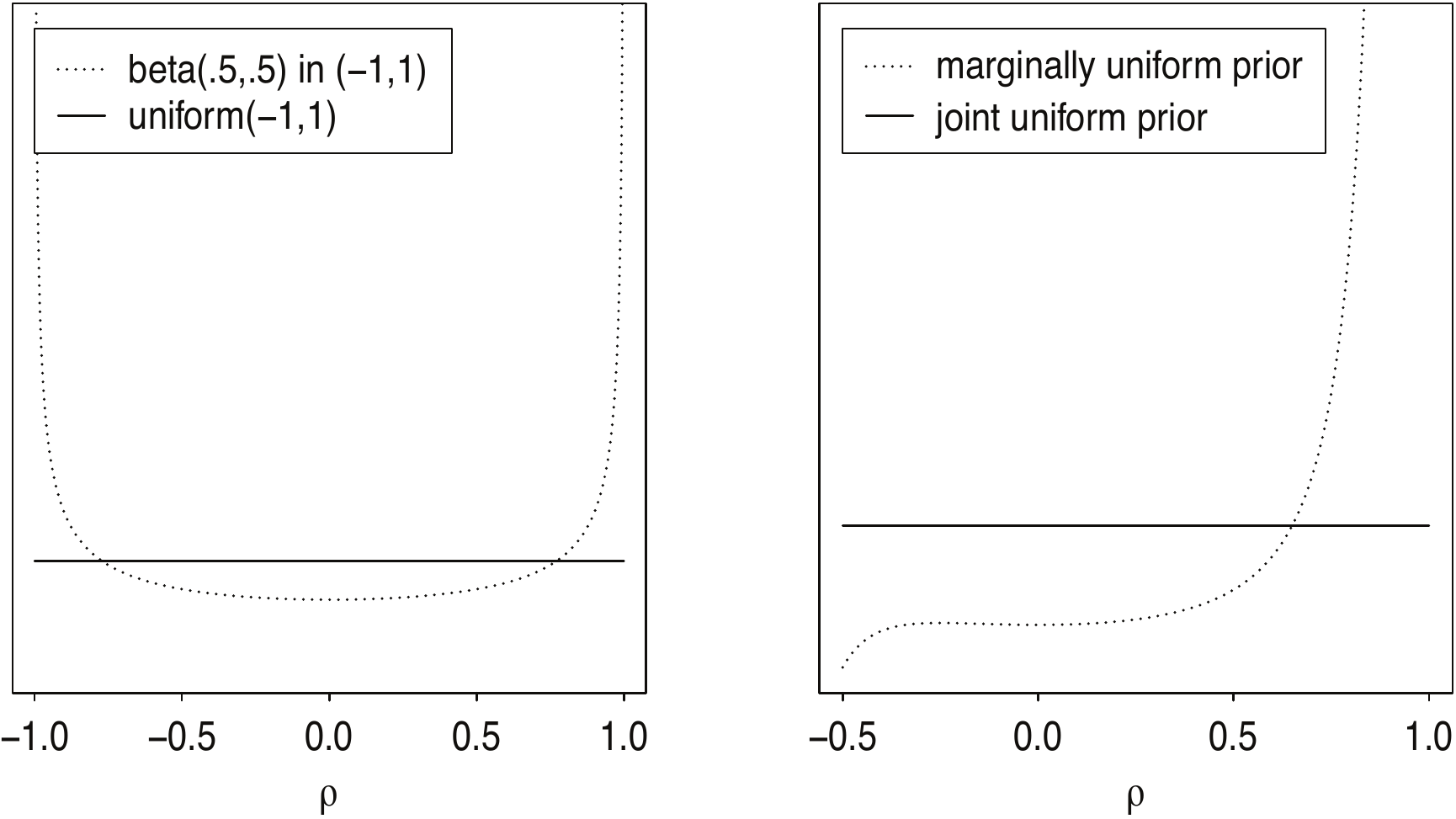}
\caption{Left panel. The implied $beta(\frac{1}{2},\frac{1}{2})$ prior in the interval $(-1,1)$ (dotted line) in the test proposed by Wetzels \& Wagenmakers' (2012), and the uniform prior in $(-1,1)$ as proposed here. Right panel. Implied prior for the common correlation $\rho$ under $H_0:\rho_{12}=\rho_{13}=\rho_{23}$ when testing against $H_1:\rho_{12}\not=\rho_{13}\not =\rho_{23}$ using a marginally uniform encompassing prior (dotted line), and a uniform prior for $\rho$ on $(-\frac{1}{2},1)$ (solid line) as proposed here.}
\label{figprior}
\end{figure}

\noindent\textit{Comparison with scale mixtures of $g$ priors}\\
\cite{Wetzels:2012} proposed a test for a single bivariate or partial correlation, $H_0:\rho=0$ against $H_1:\rho\not = 0$, via a scale mixture of $g$ priors \citep{ZellnerSiow:1980,Liang:2008,Rouder:2012b}. Their test is formulated under a linear regression model $y_i=\beta_0+\beta_1x_i+\epsilon_i$, such that the hypothesis test is equivalent to testing $H_0:\beta_1=0$ against $H_1:\beta_1\not = 0$ where $\rho$ is the correlation between $Y$ and $X$. Under the alternative hypothesis, a $g$ prior \citep{Zellner:1986} is specified for $\beta$ with an inverse gamma mixing prior for $g$. It can be shown (Appendix A) that this prior is equivalent to a $beta(\frac{1}{2},\frac{1}{2})$ prior in the interval $(-1,1)$ for $\rho$ (left panel of Figure \ref{figprior}; dotted line). As can be seen this prior puts most probability mass in the extreme regions near $-1$ and $1$. For this reason this Bayes factor will result in an overestimation of the evidence in favor of $H_0$ because it assumes unrealistically large correlations to be most plausible under the alternative hypothesis. The uniform prior for $\rho$ (left panel of Figure \ref{figprior}; solid line) on the other hand seems a better operationalization of `prior ignorance' because it assumes that all correlations under the alternative are equally likely a priori. \\

\noindent\textit{Comparison with a marginally uniform encompassing prior}\\
\cite{Barnard:2000} showed how to construct a prior for a correlation matrix having uniform marginal priors for the separate bivariate correlations. For a $P\times P$ correlation matrix, this can simply be achieved by placing an inverse Wishart prior on the covariance matrix with an identity scale matrix and $P+1$ degrees of freedom. Although the prior is very reasonable for Bayesian estimation (as shown by Barnard et al.), this marginally uniform prior distribution may not be reasonable as an encompassing prior when testing hypotheses with equality constraints on the correlations. To construct a prior that has uniformly distributed marginal priors for the separate correlations, most probability mass needs to be placed near the extremes \citep[see Figure 2 of][]{Barnard:2000}. This will result in unrealistic priors for correlations under equality constrained hypotheses. As an example consider a hypothesis test of
$H_0:\rho_{12}=\rho_{13}=\rho_{23}$ against $H_1:\rho_{12}\not=\rho_{13}\not =\rho_{23}$, and let us construct a prior for the common correlation $\rho=\rho_{12}=\rho_{13}=\rho_{23}$ under $H_0$ that is proportional to a marginally uniform encompassing prior. This yields a prior that is proportional to $\pi_0(\rho)\propto (2\rho+1)(1-\rho)^{-5/2}(\rho+1)^{-9/2}$ in the interval $(-\frac{1}{2},1)$ (Appendix A), which is plotted in the right panel of Figure \ref{figprior} (dotted line). This implied prior is concentrated near 1 which does not correspond to reasonable prior beliefs under $H_0$. Therefore a marginally uniform prior is not recommendable as default encompassing prior for testing hypotheses with equality constraints on the correlations. A uniform prior on $(-\frac{1}{2},1)$ would be a better default choice (solid line).

Note that \cite{Mulder:2016} suggested a similar prior based on $P$ degrees of freedom when testing order-constrained hypotheses and one-sided hypotheses on correlations. This results corresponds in $beta(\frac{1}{2},\frac{1}{2})$-distributed marginal priors in the interval $(-1,1)$ for the separate correlations \citep{Barnard:2000}. Even though this may seem to be an unrealistic prior (as noted earlier), the Bayes factor is quite insensitive to the prior when testing order hypotheses because the Jeffreys-Lindley-Bartlett paradox does not play a role \citep{Mulder:2014a}. The use of a vague prior based on the minimal of $P$ degrees of freedom would actually be preferably as it would result in least prior shrinkage.

\subsection{An expression of the Bayes factor for testing (in)equality constrained hypotheses}
In order to compute the marginal likelihood under each constrained hypothesis via \eqref{marglike} using the constrained uniform prior in \eqref{jointprior} a complex multivariate integral must be solved. This endeavor can be somewhat simplified by using the fact that the uniform prior for the correlations under each hypothesis $H_t$ can be written as a truncation of a uniform unconstrained prior under the unconstrained hypothesis $H_u$, i.e.,
\begin{equation}
\pi_t^U(\bm\beta,\bm\sigma,\bm\rho) = \frac{V_u}{V_t} \times \pi^U_u(\bm\beta,\bm\sigma,\bm\rho) \times I(\bm\rho\in\mathcal{C}_t)
\label{truncprior}
\end{equation}
where the unconstrained prior is given by
\begin{equation}
\pi_u^U(\bm\beta,\bm\sigma,\bm\rho)=V_u^{-1} \times \sigma_{1,1}^{-1}\times\ldots\times\sigma_{1,P_1}^{-1} \times I(\bm\rho\in\mathcal{C}_u),
\label{jointprior}
\end{equation}
where $V_u=\int_{\mathcal{C}_u} 1 d\bm\rho$, which can be interpreted as the volume of the subspace of $\bm\rho$ that results in positive definite correlation matrices under all populations. Note that the normalizing constant in \eqref{truncprior} is equal to the reciprocal of the unconstrained prior integrated over the constrained space $\mathcal{C}_t$,
\begin{equation}
\int_{\mathcal{C}_
t} \pi^U_u(\bm\rho) d\bm\rho = V_u^{-1}\int_{\mathcal{C}_t} 1 d\bm\rho = \frac{V_t}{V_u}.
\label{normprior}
\end{equation}
%Thus, the constrained prior under $H_t$ can also be written as $\pi_t^U(\bm\beta,\bm\sigma,\bm\rho) = \frac{V_u}{V_t} \times \pi^U_u(\bm\beta,\bm\sigma,\bm\rho) \times I(\bm\rho\in\mathcal{C}_t)$

This relationship between each constrained prior and the unconstrained prior allows us to use the following general result for computing the Bayes factor of a hypothesis with certain constraints on the parameters of interest against a larger unconstrained hypothesis in which the constrained hypothesis is nested.

\begin{lemma}
Consider a model where a hypothesis is formulated with equality and inequality constraints on the parameter vector $\bm\rho$ of length $Q$ of the form $H_t:\textbf{R}_t^E\bm\rho=\textbf{r}_t^E,~\textbf{R}_t^I\bm\rho>\textbf{r}_t^I$, where the $q^E\times Q$ matrix $\textbf{R}_t^E$ has rank $q^E\le Q$, and a larger `unconstrained' hypothesis $H_u$ in which $H_q$ is nested. Let the vector of nuisance parameters in the model are denoted by $\bm\phi$. If the prior of $\bm\rho$ under $H_t$ is defined as the truncation of a proper prior for $\bm\rho$ under $H_u$ in its constrained subspace, i.e., $\pi_t(\bm\rho)\propto\pi_u(\bm\rho)I(\bm\rho\in\mathcal{C}_t)$, then the Bayes factor of $H_t$ against $H_u$ can be written as
\begin{eqnarray}
B_{tu} = \frac{\mbox{Pr}(\tilde{\textbf{R}}_t^I\bm\rho^{I}_t>\textbf{r}_t^I|\bm\rho_t^E=\textbf{r}_t^E,H_u,\textbf{Y},\textbf{X})}
{\mbox{Pr}(\tilde{\textbf{R}}_t^I\bm\rho_t^{I}>\textbf{r}_t^I|\bm\rho^E_t=\textbf{r}_t^E,H_u)}\times
\frac{\pi_u(\bm\rho_t^E=\textbf{r}^E_t|\textbf{Y},\textbf{X})}{\pi_u(\bm\rho_t^E=\textbf{r}^E_t)},
\label{LemmaBFtu}
\end{eqnarray}
where $\bm\rho^E_t=\textbf{R}_t^{E}\bm\rho$, $\bm\rho_t^I$ are the elements of $\bm\rho$ that are not restricted with equality constraints under $H_t$, and $\tilde{\textbf{R}}_t^I$ consists of the columns of $\textbf{R}_t^I$ such that $\tilde{\textbf{R}}_t^I\bm\rho_t^I=\textbf{R}_t^I\bm\rho$.
\end{lemma}
\textbf{Proof.} The derivation is a combined result of \cite{Dickey:1971}, \cite{Klugkist:2005}, \cite{Pericchi:2008}, \cite{Wetzels:2010}, and \cite{Gu:2017}. A proof is given in Appendix B.\\

Note that the second factor in \eqref{LemmaBFtu} is equal to the well-known Savage-Dickey density ratio \citep{Dickey:1971,Verdinelli:1995,Wetzels:2010}. The ratio of posterior and prior probabilities in the first factor was also observed in \cite{Klugkist:2005} when there are no equality constraints under $H_t$. The conditional posterior probability, i.e., the numerator of the first term of \eqref{LemmaBFtu}, can be interpreted as a measure of fit of the order constraints of $H_t$ relative to $H_u$; the marginal posterior density, i.e., the numerator of the second term of \eqref{LemmaBFtu}, can be seen as a measure of fit of the equality constraints of $H_t$ relative to $H_u$; the conditional prior probability, i.e., the denominator of the first term of \eqref{LemmaBFtu}, can be interpreted as a measure of complexity of the order constraints of $H_t$ relative to $H_u$; the marginal prior density, i.e., the denominator of the second term of \eqref{LemmaBFtu}, can be seen as a measure of complexity based on the equality constraints of $H_t$ relative to $H_u$; see also \cite{Mulder:2014a} and \cite{Gu:2017}. Evaluating equality constraints and the inequality constraints conditional on the equality constraints separately was shown by \cite{Pericchi:2008}. The contribution here is that the Bayes factor is derived for the general case of a set of linear equality constraints and a set of linear inequality constraints where the prior under $H_t$ is a truncation of the unconstrained prior. A similar result was derived for the (adjusted) fractional Bayes factor by \cite{MulderOlsson:2019}.

In the current paper, the nuisance parameters $\bm\phi$ would correspond to the vector of the elements $\textbf{B}$ and $\bm\sigma$. Expression \eqref{LemmaBFtu} shows that the Bayes factor for a constrained hypothesis against the unconstrained hypothesis only depends on the prior for the nuisance parameters through the unconstrained marginal posterior of the correlations $\bm\rho$, i.e.,
\begin{eqnarray*}
\pi_u(\bm\rho|\textbf{Y},\textbf{X}) = \iint \pi_u(\bm\rho|\textbf{B},\bm\sigma,\textbf{Y},\textbf{X})\times \pi_u(\textbf{B},\bm\sigma|\textbf{Y},\textbf{X}) d\textbf{B} d\bm\sigma,
\end{eqnarray*}
where
\[
\pi_u(\textbf{B},\bm\sigma|\textbf{Y},\textbf{X}) \propto \int p(\textbf{Y}|\textbf{X},\textbf{B},\bm\sigma,\bm\rho)\times \pi_u^N(\textbf{B})\times \pi_u^N(\bm\sigma)\times \pi_u(\bm\rho)d\bm\rho.
\]
%
%\begin{eqnarray*}
%\pi_u(\bm\rho|\textbf{Y},\textbf{X}) = \iint \pi_u(\textbf{B},\bm\sigma,\bm\rho|\textbf{Y},\textbf{X}) d\textbf{B} d\bm\sigma,
%\end{eqnarray*}
%where
%\[
%\pi_u(\textbf{B},\bm\sigma,\bm\rho|\textbf{Y},\textbf{X}) \propto p(\textbf{Y}|\textbf{X},\textbf{B},\bm\sigma,\bm\rho)\times \pi_u^N(\textbf{B})\times \pi_u^N(\bm\sigma)\times \pi_u(\bm\rho).
%\]
%
%
As is well-known from Bayesian estimation, if different vague priors would have been specified for $\textbf{B}$ and $\bm\sigma$, e.g., a matrix-normal prior or an inverse gamma prior, respectively, the unconstrained posterior for $\textbf{B}$ and $\bm\sigma$ would have been virtually the same, and thus, the marginal posterior for $\bm\rho$ would have been virtually the same. Therefore, the Bayes factor in \eqref{LemmaBFtu} will be insensitive to the exact choice of the priors for the common nuisance parameters, as long as they are vague. This justifies the chosen noninformative independence Jeffreys priors.

Under mild circumstances, Bayes factors are known to be consistent \citep[e.g.,][]{OHagan:1995}. Loosely formulated this implies that the evidence towards the true constrained hypothesis goes to infinity as the sample size goes to infinity (this is illustrated for a specific situation in Section 6). The consistency of the proposed Bayes factor can also be observed from expression \eqref{LemmaBFtu}. In the case of a hypothesis with only equality constraints, the unconstrained posterior density (the numerator of right term) would go to infinity if the true parameter values satisfy the equality constraints as the sample size goes to infinity. If the equality constraints would not be satisfied, the posterior density would go to zero in the limit. This follows directly from large sample theory. In the case of a hypothesis with only inequality constraints the posterior probability (the numerator of the left term) would go to one if the constraints hold, and to zero if the constraints would not hold. Finally, in the case of a hypothesis with both equality and inequality constraints the product of the numerators would go to infinity if the constraints hold, and to zero elsewhere. As a result, the evidence for the true hypothesis, as quantified by the Bayes factor, would go to infinity. Consequently, the posterior probability for the true hypothesis will always go to one as the sample size grows.

\section{Numerical computation}\label{sectioncomp}
In this section we discuss a general numerical method to compute the elements in \eqref{LemmaBFtu} for a hypothesis $H_t$ with equality and/or inequality constraints on the measures of association in the generalized multivariate probit model in Section \ref{sectionmodel} using the uniform prior discussed in Section \ref{sectionprior}. The computation of the posterior parts in the numerators in \eqref{LemmaBFtu} is discussed first, followed by the prior parts in the denominators.

\subsection{Computation of the posterior probability and posterior density}
To compute the posterior probability and the posterior density in the numerators in \eqref{LemmaBFtu} under the unconstrained hypothesis, the unconstrained marginal posterior for $\bm\rho$ needs to approximated. This is done using posterior draws of the parameters from the unconstrained generalized multivariate probit model (Section \ref{sectionmodel}) using a MCMC sampler. To sample the group specific correlation matrices $\textbf{C}_g$, the Metropolis-Hastings step of \cite{Liu:2006} is extended to the generalized multivariate probit model with both continuous and ordinal outcome variables. In this sampling step we use a uniform candidate prior for the covariance matrix, i.e., $\pi(\bm\Sigma_g)\propto 1$, which is equivalent to the joint uniform target prior of the correlation matrix in \eqref{corrprior} so that each draw is always accepted. Furthermore, to ensure fast mixing of the threshold parameters, the parameter expansion method of \cite{LiuSabatti:2000} is extended to the generalized multivariate probit model with a scale group that is unique for each group and each dimension $p$, for $p=P_1,\ldots,P$. Note that \cite{Raach:2005} showed the superiority of Liu and Sabatti's (2000) parameter expansion in comparison to other sampling procedures. Details about the conditional distributions are given in Appendix C. The Fortran code of the MCMC algorithm can be found on {\tt Github/jomulder/BCT}.

Next, a Fisher $z$-transformation is applied to the unconstrained posterior sample for $\bm\rho$. Let $\rho_{p_1p_2}$ be the association between the $p_1$-th and $p_2$-th dependent variable. Then,
\[
\eta_{p_1p_2}=\tfrac{1}{2}\log\left(\tfrac{1+\rho_{p_1p_2}}{1-\rho_{p_1p_2}}\right)
\]
is the corresponding Fisher transformed measure of association. The unconstrained posterior of the Fisher transformed parameter follows an approximate normal distribution. This can be seen as follows. First note that the posterior is proportional to the likelihood times the prior. In this paper, a flat (uniform) prior is used for the correlations, and thus, the posterior is essentially proportional to the likelihood. In the integrated likelihood (where the nuisance parameters are integrated out) the sample correlation $r$ is known to have a similar role as the population correlation $\rho$\footnote{The integrated likelihood is given by \citep{JohnsonKotz:1970}:
$
f(r|\rho)=\frac{(1-\rho^2)^{(n-1)/2}(1-r^2)^{(n-4)/2}}{\pi(n-3)!}
\frac{d^{n-2}}{d(\rho r)^{n-2}}\left\{ \frac{\cos^{-1}(-\rho r)}{\sqrt{1-(\rho r)^2}} \right\}$.}. Now because the Fisher transformed sample correlation given the population correlation is also known to be approximately normal, the Fisher transformed population correlation given the sample correlation (i.e., the integrated posterior) will also be approximately normal.

\begin{figure}[t]
\centering
\makebox{\includegraphics[height=6cm]{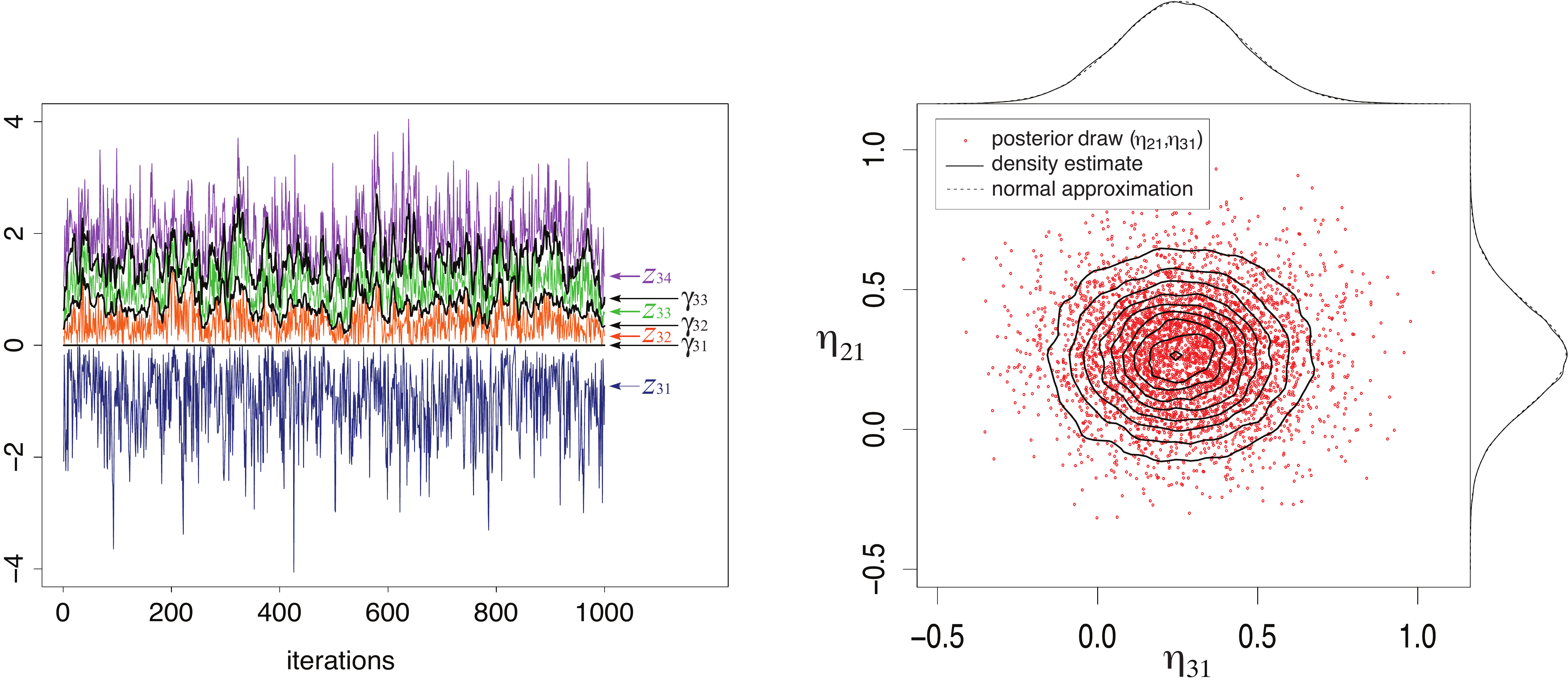}}
\caption{Left panel: Trace plot of latent $z_{31}$ observed in category 1 of the 3th (ordinal) outcome variable (blue line), $z_{32}$ observed in category 2 (orange line), $z_{33}$ observed in category 3 (green line), and $z_{34}$ observed in category 4 (purple line), as well as the corresponding threshold parameters $\gamma_{31}$, $\gamma_{32}$, and $\gamma_{34}$ for the 3th outcome variable. Right panel: Scatter plot of posterior draws of $(\eta_{21},\eta_{31})$ (red dots) with additional contour plot and univariate density plots (solid lines) and normal approximations (dashed lines).} \label{fig3}
\end{figure}

To illustrate the accuracy of the normal approximation, a sample of size $n=40$ was generated from a generalized multivariate probit model with $P=3$ dimensions where the first dependent variable was normally distributed, the second dependent variable was ordinal with two categories, and the third dependent variable was ordinal with four categories. The population variables were set to $\bm\rho'=(\rho_{21},\rho_{31},\rho_{32})=(.25,.25,0)$. A posterior sample of 10,000 draws was obtained for $\bm\rho$ and transformed to the respective Fisher transformed parameters $\bm\eta=(\eta_{21},\eta_{31},\eta_{32})$. The traceplots for the first 1,000 draws for four different latent $z$-scores belonging to observations in the four different categories of the third outcome variable together with the corresponding threshold parameters are plotted in Figure \ref{fig3} (left panel), and the posterior draws of $(\eta_{21},\eta_{31})$ are plotted in Figure \ref{fig3} (right panel) together with a contour plot of the bivariate density and the density estimates of the univariate posteriors. Normal approximations are also displayed (dashed lines). As can be seen the normal approximation is very accurate. For other settings (e.g., different dimensions or measurement levels of the outcome variables), the plots looked similar. For larger samples, the normal approximations are even better.

Thus, we can write $\pi(\bm\eta|\textbf{Y})\approx N(\bm\mu_{\bm\eta},\bm\Sigma_{\bm\eta})$, where the posterior mean $\bm\mu_{\bm\eta}$ and covariance matrix $\bm\Sigma_{\bm\eta}$ can be estimated from the posterior sample. Subsequently, the linear transformation $\bm\xi=\textbf{T}\bm\eta$ is used, with $\textbf{T}=[\textbf{R}^{E'}_t~\textbf{R}_t']$ as in Lemma 1. Approximately normality also holds for $\bm\xi$, with $\pi(\bm\xi|\textbf{Y})\approx N(\textbf{T}\bm\mu_{\bm\eta},\textbf{T}\bm\Sigma_{\bm\eta}\textbf{T}')$. Therefore, the posterior density can be estimated by plugging in $\bm\xi^E=\textbf{r}^E_t$ in the multivariate normal density $\pi(\bm\xi^E|\textbf{Y})\approx N(\textbf{R}_t^E\bm\mu_{\bm\eta},\textbf{R}_t^E\bm\Sigma_{\bm\eta}\textbf{R}_t^{E'})$. The conditional posterior probability can also be efficiently computed using the normal approximation. First the inequality constraints are rewritten via a linear transformation, e.g.,
\[
Pr(\eta_{21}<\eta_{31}<\eta_{32})=Pr(\zeta_{1}>0,\zeta_2>0)=Pr(\zeta_{1}>0|\zeta_2>0)\times Pr(\zeta_2>0)
\]
where $(\zeta_1,\zeta_2)=(\eta_{31}-\eta_{21},\eta_{32}-\eta_{31})$ \citep{Mulder:2016}. The conditional probabilities can then be estimated efficiently from MCMC output \citep{Morey:2011,Gu:2017}.

\subsection{Computation of the prior probability and prior density}
Similar as the posterior components, the conditional prior probability and prior density in \eqref{LemmaBFtu}, are computed by first approximating the unconstrained uniform prior using a prior sample. This can be done using the algorithm of \cite{Joe:2006}\footnote{The Fortran code for the sampler can be found on {\tt github.com/jomulder/BCT}}. To avoid the Borel-Kolmogorov paradox in the case of equality constraints \citep{Wetzels:2010}, the Fisher transformation is also applied to the prior draws of the $\rho$'s, resulting in prior draws for the corresponding $\eta$'s. Substantially similarly as for the posterior, a linear transformation is applied to the prior draws for $\bm\xi$, i.e., $\bm\xi=\textbf{T}\bm\eta$. Unlike the posterior, the prior of $\bm\xi$ is not approximately normal.

To estimate the prior density of $\bm\xi$ at $\tilde{\textbf{r}}_t^E$, (i.e., the Fisher transformed values of $\textbf{r}_t^E$), we use the fact that
\begin{eqnarray*}
Pr(|\xi_1-\tilde{r}_1^E|<\tfrac{\delta}{2},\ldots,|\xi_{q^E}-\tilde{r}_{q^E}^E|<\tfrac{\delta}{2}|H_u) \approx \delta^{q^E}\times \pi_u(\bm\xi^E=\tilde{\textbf{\textbf{r}}}^E)
\end{eqnarray*}
for sufficiently small $\delta$. Because a large prior sample can efficiently be obtained without needing MCMC, estimating the above probability as the proportion of draws satisfying the constraints is quite efficient. Thus, the prior density will be estimated as
\[
\hat{\pi}_u(\bm\xi^E=\tilde{\textbf{r}}^E) = \delta^{-q^E}S^{-1}\sum_{s=1}^S I(|\xi_1^{(s)}-\tilde{r}_1^E|<\tfrac{\delta}{2},\ldots,|\xi^{(s)}_{q^E}-\tilde{r}_{q^E}^E|<\tfrac{\delta}{2}),
\]
for sufficiently large $S$ and a some small value for $\delta>0$.

To compute the conditional prior probability, we approximate the prior conditional of $\bm\xi^I$ given $\bm\xi^E=\tilde{\textbf{r}}^E$ with a normal distribution where the mean and covariance matrix are estimated as the arithmetic mean and the least squares estimate based on the prior draws that satisfy $\bm\xi^E\approx\tilde{\textbf{r}}^E$, respectively. The same approximation is used as for the equality constraints, i.e., $|\xi_q^{(s)}-\tilde{r}_q^E|<\tfrac{\delta}{2}$, for $q=1,\ldots,q^E$, where $\xi_q^{(s)}$ is the $s$-th draw of $\xi_q$. A normal approximation is justified for the computation of the conditional prior probability because this probability is not very sensitive to the exact distributional form. For example, the probability that a parameter is larger than 0 is identical for a uniform distribution in $(-1,1)$ as for a standard normal distribution or for any other symmetrical distribution around zero. Note that this is not the case for the prior density at 0 which is why a normal approximation was not used for estimating the prior density. Based on the normal approximation, which can be summarized as $\pi_u(\bm\xi^E\approx\textbf{r}^E)\approx N(\bm\mu_{\bm\xi0},\bm\Sigma_{\bm\xi0})$, the same procedure can be used for estimating the conditional prior probability of $\textbf{R}_t^I\textbf{T}^{-1}[\textbf{r}^{E'},\bm\xi^{I'}]'>\textbf{r}_t^I$, as was used for the conditional posterior probability.

\section{Performance of the Bayes factor test}
\begin{figure}[t]
\centering
\makebox{\includegraphics[width=12.0cm,height=9cm]{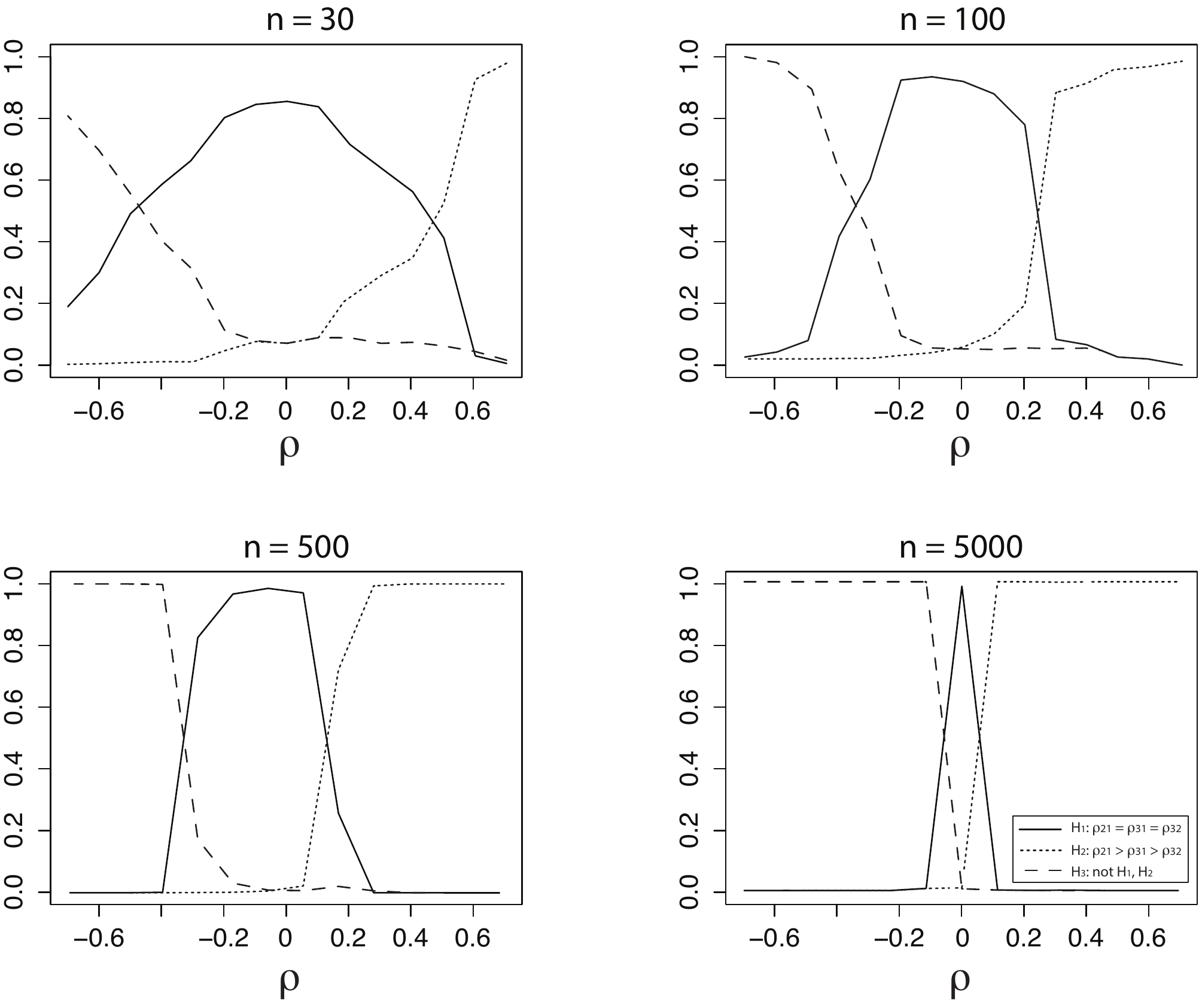}}
\caption{Posterior probabilities of $H_1:\rho_{21}=\rho_{31}=\rho_{32}$ (solid line), $H_1:\rho_{21}>\rho_{31}>\rho_{32}$ (dotted line), and $H_2:$ not $H_1,H_2$ (dashed line) for different effects $\rho$ and different sample sizes $n$.} \label{fig2}
\end{figure}

To illustrate the behavior of the methodology we consider a multiple hypothesis test of 
\begin{eqnarray*}
H_1&:&\rho_{21}=\rho_{31}=\rho_{32}\\
H_2&:&\rho_{21}>\rho_{31}>\rho_{32}\\
H_3&:&\mbox{not }H_1,H_2,
\end{eqnarray*}
for a generalized multivariate probit model with $P=3$ dependent variables of which the first is normally distributed (with variance 1), the second is ordinal with two categories, and the third is ordinal with three categories for one population (the population index $j$ is therefore omitted) and an intercept matrix $\textbf{B}=(1,1,1)'$. Data sets were generated for populations where the matrix with the measures of association were equal to
\[
\textbf{C}=\left[\begin{array}{ccc}
1\\
\rho_{21} & 1\\
\rho_{31} & \rho_{32} & 1\\
\end{array}\right]
=
\left[\begin{array}{ccc}
1\\
\rho & 1\\
\tfrac{1}{2}\rho & 0 & 1\\
\end{array}\right],
\]
for $\rho=-.7,-.6,\ldots,.6,.7$. Note that for $\rho=0$, $\rho>0$, and $\rho<0$, hypothesis $H_1$, $H_2$, and $H_3$ are true, respectively. Sample sizes of $n=30,~100,~500,$ and 5000 were considered. Equal prior probabilities were set for the hypotheses, i.e., $P(H_1)=P(H_2)=P(H_3)$. For each data set, the Bayes factor of the constrained hypotheses against the unconstrained hypothesis were first computed using the methodology of Section \ref{sectioncomp}. Note that the Bayes factor of the complement hypothesis $H_3$ against the unconstrained hypothesis can be obtained as the ratio of posterior and prior probabilities that the order constraints of $H_2$ do not hold. The equality constraints of hypothesis $H_1$ are not evaluated because they have zero probability under $H_3$. The Bayes factors between the constrained hypotheses can then be computed using its transitive relationship. Finally posterior probabilities for $H_1$, $H_2$, and $H_3$ can then be computed using \eqref{posthypo}. These posterior probabilities are plotted in Figure \ref{fig2}.

The plots show consistent behavior of the posterior probabilities for the hypotheses: as the sample size becomes very large the posterior probability of the true hypothesis goes to 1 and the posterior probabilities of the incorrect hypotheses goes to zero. Furthermore it can be seen that for small data sets with $n=30$, relatively large effects (approximately $\rho=.5$) need to be observed before either $H_2$ or $H_3$ (depending on the sign of the effect) receives more evidence than the null hypothesis. This implies that a null hypothesis with a uniform prior is better able predict the data than the alternative hypotheses in the case of moderate effects and small samples. As the sample size grows very small and zero effects can best be explained by the null hypothesis, and larger effects can best be explained by the alternative hypotheses, depending on the direction.

For other types of hypothesis tests the rate of evidence increases in a similar fashion as the sample size grows. We refer the interested reader to the increasing literature that further explores the accumulation of the evidence in the data for a true hypothesis with equality, order, or interval constraints as the sample size grows \citep[e.g.,][]{Mulder:2014b,Dittrich:2017,BoeingMessing:2017, MulderFox:2018}.

\section{Software}
The methodology is implemented in a Fortran software package to ensure fast computation and general utilization of the new methodology. The software package is referred to as `{\tt BCT}' (Bayesian Correlation Testing). The user only needs to specify the model characteristics (such as the number of dependent variables, the measurement level of the dependent variables, and the number of covariates) and the hypotheses with competing equality and order constraints on the measures of association. After running the program an output file is generated that contains the posterior probabilities of all constrained hypotheses, as well as the complement hypothesis, based on equal prior probabilities for the hypotheses. Furthermore, unconstrained Bayesian estimates of the model parameters, and 95\%-credibility intervals are provided. A user manual for {\tt BCT} can be found in Appendix E. 

\section{Exemplary hypotheses: Associations between life, leisure and relationship satisfaction (revisited)}
We now return to our empirical example and evaluate the informative hypotheses that we developed at the beginning of this contribution. Table \ref{table:PosteriorProbabilitiesEmpricalExample} reports for each set of hypotheses the posterior probabilities that each hypothesis is true after observing the data when assuming equal prior probabilities for the hypotheses.

\begin{table}
	\caption{Posterior probabilities for the competing hypotheses from Example 1 and Example 2}
	\label{table:PosteriorProbabilitiesEmpricalExample}
	\begin{tabular}{p{8	cm}p{4cm}} \hline \\
		\multicolumn{1}{l}{\textit{Example 1: Life, leisure,and relationship satisfaction}}
		\\ \hline 
		Hypotheses for men: & Posterior probabilities for the hypotheses \\ \hline 
		$H_{1_{a}}$: ${\mathrm{\rho }}_{g1y2y1}>{\mathrm{\rho }}_{g1y3y1}>{\mathrm{\rho }}_{g1y3y2}$\textit{} & 0.0020 \\  
		$H_{1_{b}}$:$\mathrm{\ }{\mathrm{\rho }}_{g1y3y1}>{\mathrm{\rho }}_{g1y2y1}>{\mathrm{\rho }}_{g1y3y2}$ & 0.0085 \\  
		$H_{1_{c}}$:$\mathrm{\ }{\mathrm{\rho }}_{g1y3y1}={\mathrm{\rho }}_{g1y2y1}={\mathrm{\rho }}_{g1y3y2}$ & 0.9879 \\ 
		$H_{1_{d}}$: not \textit{$H_{1{}_{a}}$}, nor \textit{$H_{1{}_{b}}$}, nor \textit{$H_{1{}_{c}}$} & 0.0016 \\  
		&  \\ 
		Hypotheses for women: &  \\ \hline 
		$H_{2_{a}}$: ${\mathrm{\rho }}_{g2y2y1}>{\mathrm{\rho }}_{g2y3y1}>{\mathrm{\rho }}_{g2y3y2}$ & 0.0001 \\ 
		$H_{2_{b}}$: ${\mathrm{\rho }}_{g2y3y1}>{\mathrm{\rho }}_{g2y2y1}>{\mathrm{\rho }}_{g2y3y2}$ & 0.2955 \\ 
		$H_{2_{c}}$: ${\mathrm{\rho }}_{g2y3y1}={\mathrm{\rho }}_{g2y2y1}={\mathrm{\rho }}_{g2y3y2}$ & 0.2104 \\ 
		$H_{2_{d}}$: not \textit{$H_{2{}_{a}}$,} nor \textit{$H_{2{}_{b}}$}, nor \textit{$H_{2{}_{c}}$} & 0.4940 \\  
		&  \\ 
		Hypotheses with gender as moderator: &  \\ \hline 
		$H_{3_{a}}$: ${\mathrm{\rho }}_{g\mathrm{1}y\mathrm{2}y\mathrm{1}}>{\mathrm{\rho }}_{g\mathrm{2}y\mathrm{2}y\mathrm{1}}$ & 0.1588
		\\  
		$H_{3_{b}}$: ${\mathrm{\rho }}_{g\mathrm{1}y\mathrm{2}y\mathrm{1}}={\mathrm{\rho }}_{g\mathrm{2}y\mathrm{2}y\mathrm{1}}$ & 0.8278
		\\ 
		$H_{3_{c}}$: ${\mathrm{\rho }}_{g\mathrm{1}y\mathrm{2}y\mathrm{1}}<{\mathrm{\rho }}_{g\mathrm{2}y\mathrm{2}y\mathrm{1}}$ & 0.0135
		\\  
		&  \\
		$H_{4_{a}}$: ${\mathrm{\rho }}_{g\mathrm{1}y\mathrm{3}y\mathrm{1}}>{\mathrm{\rho }}_{g\mathrm{2}y\mathrm{3}y\mathrm{1}}$ & 0.0809
		\\  
		$H_{4_{b}}$: ${\mathrm{\rho }}_{g\mathrm{1}y\mathrm{3}y\mathrm{1}}={\mathrm{\rho }}_{g\mathrm{2}y\mathrm{3}y\mathrm{1}}$ &  0.9017 \\ 
		$H_{4_{c}}$: ${\mathrm{\rho }}_{g\mathrm{1}y\mathrm{3}y\mathrm{1}}<{\mathrm{\rho }}_{g\mathrm{2}y\mathrm{3}y\mathrm{1}}$ & 0.0174
		\\ \hline \\
		\multicolumn{1}{l}{\textit{Example 2: Support for justice principles}}
		\\ \hline 
		$H_{5_{a}}$: ${\rho _{g1y2y1}} > {\rho _{g2y2y1}} > {\rho _{g3y2y1}} > {\rho _{g4y2y1}}$ &   0.0044 
		\\  
		$H_{5_{b}}$: ${\rho _{g1y2y1}} = {\rho _{g2y2y1}} = {\rho _{g3y2y1}} = {\rho _{g4y2y1}}$ &   0.0000 
		\\
		$H_{5_{c}}$: ${\rho _{g1y2y1}} = {\rho _{g2y2y1}} > {\rho _{g3y2y1}} = {\rho _{g4y2y1}}$ &   0.9899 
		\\
		$H_{5_{d}}$: not $H_{5_{a}}$ nor $H_{5_{b}}$, nor $H_{5_{c}}$  &   0.0057 \\
		\hline 
	\end{tabular}
\end{table}

Based on these results we can conclude that for men there is overwhelming evidence of equal partial associations between life satisfaction, leisure satisfaction, and relationship satisfaction, controlling for differences in mood at survey and self-reported health (99\%). 
However, for women the evidence indicates relatively large support for the complement hypothesis (49\%) and substantially weaker support for the ordered partial correlations hypothesis $H_{2_b}$ and the equal partial correlations hypothesis $H_{2_c}$.

Next, we consider the hypotheses that expect either ordered or equal conditional (i.e. with gender as a moderator) partial associations between the satisfaction variables. Here, we see that there is strong evidence for the hypotheses that life satisfaction and each type of domain-specific satisfaction relate equally strong to each other for men and women (83\% and 90\% respectively). In summary, the analysis portrays a picture of a large degree of equality of association between satisfaction indicators while holding constant for other variables, and little evidence for the expected ordering of partial associations among the variables considered.

Table \ref{table:PosteriorProbabilitiesEmpricalExample} also reports the findings for the test of the informative hypotheses on the partial correlations between justice principles. Here, the findings indicate very strong evidence for Hypothesis $H_{5_c}$. This equality and order-restricted hypothesis assumed that within the post-communist country cluster and within the Western European country cluster the partial correlation between support for both justice principles would be equal, but between country clusters the size of the partial correlation would be different, with an expected larger partial association in the post-communist group of countries.

\section{Discussion}
This paper presented a flexible framework for testing statistical hypotheses on most commonly observed measure of association in social research. By developing powerful, flexible and user-friendly Bayesian methods for testing informative hypotheses about partial correlations, we underscore the importance and appropriateness of the partial correlation coefficient as a tool for testing hypotheses about associations between variables of various measurement levels when applying regression analysis would actually not be appropriate from a research design or substantive perspective. The methodology has the following useful properties.
\begin{itemize}
\item A broad class of hypotheses can be tested with competing equality and/or order constraints on the measures of association (Table 1).
\item Multiple (more than two) hypothese can be tested simultaneously in a straightforward manner.
\item Constrained hypotheses can be formulated on tetrachoric correlations, polychoric correlations, biserial correlations, polyserial correlations, and product-moment correlations (Table 2). These measures of association can be corrected for certain covariates to avoid spurious relations.
\item A simple answer is provided to the research question which hypothesis receives most evidence from the data and how much, using Bayes factors and posterior probabilities.
\item The proposed test is consistent which implies that the posterior probability of the true hypothesis goes to one as the sample size goes to infinity.
\item The software package {\tt BCT} allows social science researchers to easily apply the methodology in real-life examples.
\end{itemize}

The proposed methodology relies on the generalized multivariate probit model for continuous and ordinal outcome variables. 
This model is very well researched in the Bayesian literature (e.g., Albert \& Chib, 1995; Chib \& Greenberg, 1998; Chen \& Dey, 2000; Barnard et al., 2000; Fox, 2005; Boscardin et al., 2008), and implemented in professional software packages such as Mplus (Asparouhov \& Muth\'{e}n, 2010) or Stata \citep{Stata}. In the case of severe violations of the distributional assumptions (i.e., normality for continuous outcome variables and a normal latent variable for ordinal outcome variables), however, the estimated unconstrained posterior of the measures of association may not be accurate. To get a better understanding of the robustness of the method, a thorough numerical simulation study on this model would be useful. Potentially the method can also become more robust to violations of normality or the assumed probit model by adopting the rank likelihood approach of \cite{Hoff:2007}. This will be an interesting future extension yielding a more accurate Bayes factor for testing measures of association in the case of severe model violations. 

The proposed methodology was designed as a confirmatory criterion for testing multiple hypotheses with competing equality and/or order constraints on the measures of association of interest. The confirmatory aspect justifies the default use of equal prior probabilities of the hypotheses that are formulated. The method can also be used for exploratory testing of all possible hypotheses with combinations of equality and order constraints. In this case a correction for multiple testing is necessary because the number of hypotheses can become quite extensive. In a Bayesian framework such a correction can be incorporated through the prior probabilities of the hypotheses. \cite{ScottBerger:2006} showed how this can be done when exploratory testing many precise hypotheses. How this can be done when exploratory testing all possible equality and order hypotheses is still an open problem worthy of further research.

Finally the proposed Bayes factor test is based on uniform priors on the measures of association under the hypotheses of interest. This choice does not allow users to manually specify priors based on external prior beliefs about the measures of association. Although this may be viewed as a limitation, from a default `noninformative' Bayesian perspective, the class of uniform priors seems the only justifiable choice because uniform priors imply that all possible values of the parameters of interest are equally likely before observing the data. Furthermore, the proposed methodology is quite flexible as it allows researchers to formulate very specific hypotheses with equality and order constraints on the measures of association (Table \ref{tablehyp}). In fact by formulating hypotheses with very specific sets of constraints, very informative priors are implicitly specified. For example, when considering a hypothesis with equal correlations, $H_1:\rho_{12}=\rho_{13}=\rho_{23}$, the underlying prior is only positive (and constant) where all correlations are exactly equal. Similarly, the precise hypothesis $H_0:\rho=0$ corresponds to an extremely informative prior which places all its mass where $\rho$ equals 0. Thus instead of allowing users to incorporate external information by directly specifying informative priors under the hypotheses, the methodology allows users to formulate very specific hypotheses which indirectly correspond to very informative priors. In our experience translating prior beliefs to constrained hypotheses is generally easier (and less controversial) than translating prior beliefs to informative priors on the parameters themselves. The methodology is therefore suitable for testing competing scientific expectations on measures of association in a default Bayesian manner.

\subsection*{Acknowledgements}
The authors wishes to thank Andrew Tomarken for helping with testing and debugging the `BCT' software program. The first author was supported by a Veni Grant (016.145.205) provided by The Netherlands Organization for Scientific Research (NWO).

\appendix

\section{Derivation of the priors}
\textit{Implied prior by \cite{Wetzels:2012}}\\
Assume we are interested in the correlation $\rho$ between $Y$ and $X$ having a bivariate normal distribution,
\[
\left[
\begin{array}{c}Y\\ X\end{array}
\right]
\sim N\left(\left[
\begin{array}{c}\mu_Y\\ \mu_X\end{array}
\right],
\left[
\begin{array}{cc}\sigma^2_Y & \rho\sigma_{Y}\sigma_{X}\\ \rho\sigma_{Y}\sigma_{X} &  \sigma_{X}^2 \end{array}
\right]\right),
\]
where $\rho$ is the correlation between $Y$ and $X$. We assume that the variable $X$ is normalized, i.e., $\mu_X=0$ and $\sigma^2_X=1$. The bivariate normal model correspond with the following conditional formulation,
\begin{equation}
Y|X\sim N(\mu_Y+\rho\sigma_{Y}X,\sigma^2_Y(1-\rho^2)).
\label{regr2}
\end{equation}
Now we consider an alternative parameterization using a linear regression model
\begin{equation}
Y|X \sim N(\beta_0+\beta_1X,\sigma^2_{\epsilon})
\label{regr1}
\end{equation}
where $\sigma^2_{\epsilon}$ is the error variance in the regression model.
When linking these two parameterizations, this implies
\[
\beta_1=\rho\sigma_Y=\rho(1-\rho^2)^{-1}\sigma_{\epsilon} \Leftrightarrow \rho=\frac{\beta_1}{\sqrt{\beta_1^2+\sigma_{\epsilon}^2}}.
\]
Hence, testing $H_0:\rho=0$ against $H_1:\rho\not=0$ in \eqref{regr2} is equivalent to testing $H_0:\beta_1=0$ against $H_1:\beta_1\not=0$ in \eqref{regr1}.

\cite{Wetzels:2012} considered a $g$ prior for $\beta_1$ with an inverse gamma with a shape parameter of $\frac{1}{2}$ and a scale parameter of $\frac{n}{2}$, which is equivalent to a Student $t$ prior with zero location, a scale of $\sigma^2_{\epsilon}$, and 1 degree of freedom (i.e., a Cauchy prior):
\begin{eqnarray*}
\pi_1(\beta_1|\sigma^2_{\epsilon}) &=& \int N(\beta_1;0,g\sigma^2_{\epsilon}(\textbf{x}'\textbf{x})^{-1})IG(g;\tfrac{1}{2},\tfrac{n}{2})dg\\
&=& t(\beta_1;0,\sigma^2_{\epsilon},1)\\
&\propto & (1+\beta_1^2/\sigma^2_{\epsilon})^{-1},
\end{eqnarray*}
where we set $\textbf{x}'\textbf{x}=n$.
When noting that the Jacobian equals $\frac{d\beta_1}{d\rho}=\sigma_{\epsilon}(\rho^2(1-\rho^2)^{-3/2}+(1-\rho^2)^{-1/2})$, the prior for $\rho$ can be obtained by applying standard calculus, i.e.,
\begin{eqnarray*}
\pi_1(\rho|\sigma_{\epsilon}^2) &=& \pi_1(\beta_1=\rho(1-\rho^2)^{-1}\sigma_{\epsilon}|\sigma_{\epsilon}^2)\frac{d\beta_1}{d\rho}\\
&\propto&\left(1+\rho^2(1-\rho^2)^{-1}\right)^{-1}(\rho^2(1-\rho^2)^{-3/2}+(1-\rho^2)^{-1/2})\\
&= & (1-\rho^2)^{-1/2}\\
&\propto & beta(\tfrac{1}{2},\tfrac{1}{2}) \text{ in the interval }(-1,1). 
\end{eqnarray*}

Similarly, it can be shown that a uniform prior for $\rho$ in the interval $(-1,1)$ would correspond to conditional prior for $\beta_1$ given $\sigma^2_{\epsilon}$ with a Student $t$ distribution with location, scale, and degrees of freedom equal to 0, $\sigma^2_{\epsilon}/2$, and 2, respectively.\\ \smallskip\\

\noindent\textit{Marginally uniform prior approach}\\
As shown by \cite{Barnard:2000} an inverse Wishart prior with identity scale matrix $\textbf{P}$ and $P$ degrees of freedom for a covariance matrix implies a marginal prior for a correlation matrix $\textbf{C}$ having a density
\[
\pi(\textbf{C})\propto |\textbf{C}|^{\frac{(P-1)^2}{2}-1}\prod_{p=1}^P|\textbf{C}_{pp}|^{-P/2},
\]
where $\textbf{C}_{pp}$ is the $p$-th principle submatrix of $\textbf{C}$. Now consider a hypothesis where all correlations are assumed to be equal (which implies a compound symmetry correlation structure), $H_0:\rho_{21}=\ldots=\rho_{P,P-1}$. The determinants are then a function of the common correlation $\rho$ and given by $|\textbf{C}|=((P-1)\rho+1)(1-\rho)^{P-1}$ and $|\textbf{C}_{pp}|=((P-2)\rho+1)(1-\rho)^{P-2}$. Consequently, the implied prior for $\rho$ under $H_0$ is given by
\[
\pi_0(\rho)\propto (\rho(P-1)+1)^{\frac{(P-1)^2}{2}-1}(1-\rho)^{-\frac{P^2-P-1}{2}}(\rho(P-2)+1)^{-\frac{P^2}{2}},
\]
which, for $P=3$, equals
\[
\pi_0(\rho)\propto (2\rho+1)(1-\rho)^{-5/2}(\rho+1)^{-9/2}.
\]

\section{Proof of Lemma 1}
The proof is based on the results of \cite{Klugkist:2005}, \cite{Pericchi:2008}, \cite{Mulder:2010}, \cite{Wetzels:2010}, \cite{Mulder:2014b}, and \cite{MulderOlsson:2019}. The constrained hypothesis $H_t$ is given by $H_t:\textbf{R}^E\bm\rho=\textbf{r}^E,~\textbf{R}^I\bm\rho>\textbf{r}^I$, where $[\textbf{R}^E|\textbf{r}^E]$ is a $q^E\times (Q+1)$ augmented matrix representing the equality constraints on $\bm\theta$, with $q^E\le Q$, and $[\textbf{R}^I|\textbf{r}^I]$ is a $q^I\times (Q+1)$ augmented matrix representing the inequality constraints on $\bm\rho$.

Under $H_t$ there are $q^E$ equality constraints active on $\bm\rho$, which will be denoted by $\bm\rho_t^I=\textbf{R}_t^I\bm\rho$. Therefore there are $Q-q^E$ free parameters under $H_t$ (excluding the nuisance parameters), which we shall denote by $\bm\rho^I_t$. Without loss of generality we can permute the elements of $\bm\rho$, such that we can write
\[
\left[
\begin{array}{c}
\bm\rho_t^E\\
\bm\rho_t^I
\end{array}
\right]=
\left[
\begin{array}{c}
\textbf{R}_t^E\\
\textbf{0}~\textbf{I}_{Q-q^E}
\end{array}
\right]\bm\rho.
\]
Consequently, the constrained hypothesis can equivalently be written as $H_t:\bm\rho_t^E=\textbf{r}_t^E,~\tilde{\textbf{R}}_t^I\bm\rho_t^I>\textbf{r}_t^I$, where $\tilde{\textbf{R}}_t^I$ consists of the last $Q-q^E$ columns of $\textbf{R}_t^I$.

The prior under $H_t$ is a truncation of the unconstrained prior under $H_u$ in the constrained space of $H_t$, i.e.,
\begin{eqnarray*}
\pi_t(\bm\rho_t^I)&=&c_t^{-1}\pi_u(\bm\rho_t^E=\textbf{r}_t^E,\bm\rho_t^I)I(\tilde{\textbf{R}}_t^I\bm\rho_t^I>\textbf{r}_t^I),~\text{where}\\
%&=&c_t^{-1}\pi_u(\bm\rho_t^I|\bm\rho_t^E=\textbf{r}_t^E)\pi_u(\bm\rho_t^E=\textbf{r}_t^E)I(\tilde{\textbf{R}}_t^I\bm\rho_t^I=\textbf{r}_t^I),~\text{where}
c_t &=& \text{Pr}(\tilde{\textbf{R}}_t^I\bm\rho_t^I>\textbf{r}_t^I|\bm\rho_t^E=\textbf{r}_t^E,H_u)~\pi_u(\bm\rho_t^E=\textbf{r}_t^E).
\end{eqnarray*}
Furthermore, the nuisance parameters have equal priors under both the constrained and unconstrained hypothesis, i.e., $\pi_t(\bm\phi)=\pi_u(\bm\phi)$, and the likelihood of the data under $H_t$ is a truncation of the unconstrained likelihood, i.e., $p_t(\textbf{Y}|\textbf{X},\bm\rho_t^I,\bm\phi)=p_u(\textbf{Y}|\textbf{X},\bm\rho_t^E=\textbf{r}_t^E,\bm\rho_t^I,\bm\phi)I(\tilde{\textbf{R}}_t^I\bm\rho_t^I>\textbf{r}_t^I)$, where $\textbf{Y}$ contains all outcome variables and $\textbf{X}$ contains all covariates. The Bayes factor can then be written as
\begin{eqnarray*}
B_{tu} &=& \frac{\iint_{\tilde{\textbf{R}}_t^I\bm\rho_t^I>\textbf{r}_t^I} \pi_t(\bm\rho_t^I)\pi_t(\bm\phi)p_t(\textbf{Y}|\textbf{X},\bm\rho_t^I,\bm\phi)d\bm\rho_t^I d\bm\phi}
{\iiint \pi_u(\bm\rho_t^E,\bm\rho_t^I)\pi_u(\bm\phi)p_u(\textbf{Y}|\textbf{X},\bm\rho_t^E,\bm\rho_t^I,\bm\phi)d\bm\rho_t^Ed\bm\rho_t^Id\bm\rho_t^I d\bm\phi}\\
&=&\frac{\iint_{\tilde{\textbf{R}}_t^I\bm\rho_t^I>\textbf{r}_t^I} c_t^{-1}\pi_u(\bm\rho_t^E=\textbf{r}_t^E,\bm\rho_t^I)\pi_u(\bm\phi)
p_u(\textbf{Y}|\textbf{X},\bm\rho_t^E=\textbf{r}_t^E,\bm\rho_t^I,\bm\phi)d\bm\rho_t^I d\bm\phi}
{\iiint \pi_u(\bm\rho_t^E,\bm\rho_t^I)\pi_u(\bm\phi)p_u(\textbf{Y}|\textbf{X},\bm\rho_t^E,\bm\rho_t^I,\bm\phi)d\bm\rho_t^Ed\bm\rho_t^Id\bm\rho_t^I d\bm\phi}\\
&=& c_t^{-1}\iint_{\tilde{\textbf{R}}_t^I\bm\rho_t^I>\textbf{r}_t^I}
\frac{ \pi_u(\bm\rho_t^E=\textbf{r}_t^E,\bm\rho_t^I)\pi_u(\bm\phi)
p_u(\textbf{Y}|\textbf{X},\bm\rho_t^E=\textbf{r}_t^E,\bm\rho_t^I,\bm\phi)}
{\iiint \pi_u(\bm\rho_t^E,\bm\rho_t^I)\pi_u(\bm\phi)p_u(\textbf{Y}|\textbf{X},\bm\rho_t^E,\bm\rho_t^I,\bm\phi)d\bm\rho_t^Ed\bm\rho_t^Id\bm\rho_t^I d\bm\phi}
d\bm\rho_t^I d\bm\phi\\
&=& c_t^{-1}\iint_{\tilde{\textbf{R}}_t^I\bm\rho_t^I>\textbf{r}_t^I}
\pi_u(\bm\rho_t^E=\textbf{r}_t^E,\bm\rho_t^I,\bm\phi|\textbf{Y},\textbf{X})d\bm\rho_t^I d\bm\phi\\
&=& \frac{\text{Pr}(\tilde{\textbf{R}}_t^I\bm\rho_t^I>\textbf{r}_t^I|
\bm\rho_t^E=\textbf{r}_t^E,\textbf{Y},\textbf{X},H_u)\pi_u(\bm\rho_t^E=\textbf{r}_t^E|\textbf{Y},\textbf{X})}
{\text{Pr}(\tilde{\textbf{R}}_t^I\bm\rho_t^I>\textbf{r}_t^I|
\bm\rho_t^E=\textbf{r}_t^E,H_u)\pi_u(\bm\rho_t^E=\textbf{r}_t^E)},
\end{eqnarray*}
which completes the proof.

\section{Conditional distributions for the MCMC sampler}
For $g=1,\ldots,G$, the group specific parameters are sampled in the posterior as follows.
\begin{enumerate}
\item Sample $\textbf{B}_g|\textbf{V}_g,\textbf{U}_g,\textbf{Z}_g,\textbf{X}_g,\bm\sigma_g,\textbf{C}_g\sim N_{Q\times P}(\hat{\textbf{B}}_g,(\textbf{X}_g'\textbf{X}_g)^{-1},\bm\Sigma_g)$, where $\hat{\textbf{B}}_g=(\textbf{X}_g'\textbf{X}_g)^{-1}\textbf{X}_g'\textbf{Y}^*_g$, $\textbf{Y}^*_g=[\textbf{V}_g~\textbf{Z}_g]$, i.e., a stacked matrix of $(\textbf{v}_{ig}',\textbf{z}_{ig}')$, for $i=1,\ldots,n_g$, and $\bm\Sigma_g=\text{diag}(\bm\sigma_g',\textbf{1}'_{P_2})\textbf{C}_g\text{diag}(\bm\sigma_g',\textbf{1}'_{P_2})$;
\item Sample $\textbf{C}_g$ given $\textbf{B}_g$, $\textbf{V}_g$, $\textbf{U}_g$, $\textbf{Z}_g$, $\textbf{X}_g$, $\bm\sigma_g$ using a parameter expansion of \cite{Liu:2006}. Let $\textbf{E}_g = \textbf{Y}^*_g - \textbf{X}_g\textbf{B}_g$, and the normalization of the columns $\tilde{\textbf{E}}_g=\textbf{E}_g\textbf{D}_g$, where $\sum_{i=1}^{n_g}\tilde{e}^2_{gi}=1$. The positive definite scale is then given by $\textbf{S}_g=\text{diag}(1/\bm\sigma_g',\textbf{1}'_{P_2})\tilde{\textbf{E}}_g'\tilde{\textbf{E}}_g\text{diag}(1/\bm\sigma_g',\textbf{1}'_{P_2})$. Furthermore a uniform candidate prior is considered for the covariance matrix, i.e., $\pi(\bm\Sigma_g)\propto 1$, such that the candidate covariance matrix can be drawn from an inverse Wishart distribution, $IW(n_g-P-1,\textbf{S}_g)$, from which the candidate draw for $\textbf{C}_g$ can be obtained which is always accepted because the target prior for the correlation matrix is the same\footnote{The uniform candidate prior, $\pi(\bm\Sigma_g)\propto 1$, is equivalent to a joint uniform prior on $\textbf{C}_g$ because the Jacobian of the transformation, $\bm\Sigma_g\rightarrow (\bm\sigma_g,\textbf{C}_g)$, does not depend on $\textbf{C}_g$.}.
\item The threshold parameter $\gamma_{gpk}$, for $p=P_1,\ldots,P$, and $k=2,\ldots,K_p-1$, is sampled from a uniform distribution with lowerbound being the largest $z_{gip}$ that falls in the category $k-1$, and upperbound being the smallest $z_{gip}$ that falls in the category $k$.
\item The population standard deviation $\sigma_p$, for $p=1,\ldots,P_1$, is sampled using a random walk centered around the previous draw \citep[see also][]{Liu:2006}.
\item The additional parameters due to the parameter extension of \cite{LiuSabatti:2000} are sampled using a random walk centered around the previous draw. The following scale group is chosen, $\Gamma_{gp}=\{h_{gp}>0:h_{gp}(\textbf{z}_{g p},\textbf{B}_{\cdot p},\gamma_{gp2},\ldots,\gamma_{gp(K_p-1)})=(h_{gp}\textbf{z}_{g p},h_{gp}\textbf{B}_{\cdot p},h_{gp}\gamma_{gp2},\ldots,h_{gp}\gamma_{gp(K_p-1)})\}$, for group $g$ and dimension $p$, for $p=P_1,\ldots,P$. The unimodular  Haar measure for $\Gamma_{gp}$ is $L(dh_{gp})=h_{gp}^{-1}dh_{gp}$. To sample $h_{gp}$ using the random walk, the kernel for $h_{gp}$ is of the form $h_{gp}^{n_g+Q+K_p-3}\exp(-a_{gp}h_{gp}^2-b_{gp}h_{gp})$.
\end{enumerate}
The Fortran code for the MCMC algorithm can be found on {\tt github.com/jomulder/BCT}.

\section{User manual for BCT}
The software program BCT (Bayesian Correlation Testing) can be downloaded from \textsf{www.jorismulder.com}. The folder consist of six text files, i.e., {\tt BCT\_input.txt}, {\tt BCT\_output.txt}, {\tt BCT\_output\_relComp.txt}, \\{\tt BCT\_output\_relFit.txt}, {\tt BCT\_estimates.txt}, and {\tt data.txt}, and an executable file {\tt BCT.exe}. BCT can be run by double clicking `BCT.exe'. This manual describes how to specify the input and data files and how to read the output file.

\subsection*{Parameterization}
The hypotheses are tested under the generalized multivariate probit model which assumes multivariate normal distribution for the continuous dependent variables and multivariate probit distribution (with a multivariate normal distribution for the latent variables) for the ordinal dependent variables. Multiple populations can be considered having population specific intercepts, regression coefficients, variances and measures of associations.

\subsection*{Input file}
This input file `\textsf{BCT\_input.txt}' has the following layout.

\begin{verbatim}
Input 1: model & data
#DV, #covs, intercept, #populations, Ntotal, header
3 2 1 1 50 0

Which DVs are ordinal (0=continuous, 1=ordinal)
0 1 1

Input 2: hypotheses
#hypotheses
2

#equalities, #inequalities per hypothesis
1 1
1 2

Input 3: constraints in hypotheses
Equalities H1; Inequalities H1; Equalities H2; Inequalities H2; etc.
1 2 1 1 3 1
1 3 1 1 3 2

1 3 2 0 1 0
1 3 1 1 3 2
1 2 1 1 3 1

Input 4: implementation details
seed, #draws prior, #draws posterior, #draws per constraint
123 20000 10000 5000
\end{verbatim}

Under `Input 1' the properties of the model have to specified that is used for analyzing the data. In the above example the model has 3 dependent variables ({\tt \#DV=3}), 2 covariates ({\tt \#covs=2}), an intercept should be included in the model ({\tt \#intercept=1}), is used for modeling one population ({\tt \#population=1}), and the total sample size is 50 ({\tt \#Ntotal=3}). Next it is specified that the first dependent variable is continuous and the second and third dependent variables are ordinal. Finally it is specified whether the data file has a header. In this case there is no header ({\tt header=0}). If there would be a header the first row of the data file would not be read.

Under `Input 2' the number of hypotheses under investigation are specified, which is 2 in this case ({\tt \#hypotheses=2}). Below it is specified how many equality and inequality (or order) constraints each hypothesis consists of. In the above situation the first hypothesis has 1 equality constraints and 1 inequality constraint on the correlations and the second hypothesis has 1 equality constraint and 2 inequality constraints on the correlations.

Under `Input 3' the equality and inequality constraints are specified using specific coding. Every line corresponds to an equality or inequality constraint. An equality constraint of the form $\rho_{j_1p_1p_2}=\rho_{j_2p_3p_4}$, i.e., the association between dependent variables $p_1$ and $p_2$ in population $j_1$ is equal to the association between dependent variables $p_3$ and $p_4$ in population $j_2$, is coded as ``$j_1~p_1~p_2~j_2~p_3~p_4$''. The inequality constraint $\rho_{j_1p_1p_2}>\rho_{j_2p_3p_4}$ is also coded as ``$j_1~p_1~p_2~j_2~p_3~p_4$''. It follows automatically from the specification in Input 2 whether an equality or inequality constraint is considered. Furthermore, an equality constraint of the form $\rho_{j_1p_1p_2}=d$, i.e., the association between dependent variables $p_1$ and $p_2$ in population $j_1$ is equal to $d$, is coded as ``$j_1~p_1~p_2~0~1~d$''. The inequality constraints $\rho_{j_1p_1p_2}>d$ and $\rho_{j_1p_1p_2}<d$ are coded as ``$j_1~p_1~p_2~0~1~d$'' and ``$j_1~p_1~p_2~0~-1~d$'', respectively. Note that in the current case with 2 covariates, the constraints are formulated on the partial measures of association conditional on the covariates.

Because hypothesis one consists of one equality constraint and one inequality constraint, the first line specifies the equality constraint and the second line specifies the inequality constraint. The first line ``{\tt 1 2 1 1 3 1}'' specifies the equality constraint that the correlation in population 1 between dependent variable 2 and dependent variable 1 is equal than the correlation in population 1 between variable 3 and 1. Note that the code ``{\tt 1 1 3 1 2 1}'' would have resulted in exactly the same equality constraint. The second line ``{\tt 1 3 1 1 3 2}'' specifies the inequality constraint that the correlation in population 1 between dependent variable 3 and 1 is larger than the correlation in population 1 between dependent variable 3 and 2. Next, the constraints of the second hypothesis are specified. The first line ``{\tt 1 3 2 0 1 0}'' specifies the equality constraint of the second hypothesis which states that the association between variable 3 and 2 in population 1 is equal to 0. The second line ``{\tt 1 3 1 1 3 2}'' states that the association between variable 3 and 1 is larger than the association between variable 3 and 2 both in population 1. The third line ``{\tt 1 2 1 1 3 1}'' states that the association between variable 2 and 1 is larger than the association between variable 3 and 1 again both in population 1.

Under `Input 4' some computational details need to specified. First, the seed number must be specified, which equals 123 in the above setting. Second, the number prior is specified which is equal to 20,000 in this case. Third, the number of posterior draws is specified which is equal to 10,000 in this case. Fourth, the number of draws to evaluate each inequality constraint to compute the relative complexity is set which is equal to 5,000 in this case.

\subsection*{Data file}
For the above input file the data file should look like this (only the first 6 rows are displayed).
\begin{verbatim}
-0.9686  1  5   2  -0.54   1
 0.1112  2  6   3  -0.56   1
-0.0018  1  2   5   1.12   1
-0.4265  1  2   1  -2.52   1
-1.3845  1  3   3   3.32   1
-0.8355  2  2   0   1.67   1
...
\end{verbatim}
The first row specifies that the first observation has -.9686 as outcome for the first continuous dependent variable (first column), falls in category 1 of the second dependent ordinal variable (second column), falls in category 5 of the third dependent ordinal variable (third column), has an outcome of 2 on the first covariate (fourth column), has an outcome of -.54 on the second covariate (fifth column), and belongs to population 1 (sixth column). In the case of ordinal variables, the lowest category should be 1, followed by 2, etc. If there is only one population, and thus, all observations come from the same population, the last column can be omitted as well.

\subsection*{Output files}
In the output files of BCT are {\tt BCT\_output.txt}, {\tt BCT\_output\_relComp.txt}, \\{\tt BCT\_output\_relFit.txt}, {\tt BCT\_estimates.txt}. First the file {\tt BCT\_output.txt} contains the posterior probabilities for the hypotheses when assuming equal prior probabilities. The following output was obtained from the analysis with the above input file with two specified constrained hypotheses.

\begin{verbatim}
Posterior probabilities for the hypotheses
 
Hypothesis  1
0.0001
 
Hypothesis  2
0.9973
 
Complement hypothesis*
0.0026
\end{verbatim}

Thus, the posterior probability that the first, second, and the complement hypotheses are true after observing the data equal .0001, .9973, and .0026. Thus, there is most evidence that the second hypothesis is true. Note that Bayes factors between pairs of hypotheses can be computed using these posterior probabilities because equal prior probabilities are used for the hypotheses. Further note that the complement hypothesis should only be used for inference when there are not multiple hypotheses under consideration with only inequality constraints that are nested in one another. Thus, when testing nested inequality constrained hypotheses, e.g., $H_1:\rho_{121}>\rho_{131}>\rho_{132}$ and $H_1:\rho_{121}>(\rho_{131},\rho_{132})$, where $H_1$ is nested in $H_2$, the outcome of the complement hypothesis should be ignored. Note that in the current setting the outcome for the complement hypothesis can be included when making inferences.

Second, the file {\tt BCT\_output\_relComp.txt} contains the outcomes of the relative complexity of the hypotheses. In this example the file was equal to

\begin{verbatim}
rc      rcE     rcI

Hypothesis  1
0.17010 0.42100 0.40404

Hypothesis  2
0.00405 0.64325 0.00630

Complement hypothesis*
1.00000 1.00000 1.00000
\end{verbatim}

In each row the value on the right is the relative complexity of a hypothesis based on its inequality constraints, relative to the unconstrained hypothesis. The middle value in each row is the relative complexity based on the equality constraints. And the left most value is the relative complexity of a hypothesis which is a product of the other two values. Note that for the complement hypothesis all values equal 1 because the analysis for the complement hypothesis is identical to the analysis of the unconstrained hypothesis in this setting. Therefore the relative complexities of the complement hypothesis relative to the unconstrained hypothesis equal 1.

The results in the output file {\tt BCT\_output\_relFit.txt} can be interpreted in a similar manner but then for the relative complexity of the hypotheses. For completeness, for this analysis the file looks as follows.

\begin{verbatim}
rf      rfE     rfI
 
Hypothesis  1
0.00853 0.00880 0.96903
 
Hypothesis  2
1.56453 1.89166 0.82707
 
Complement hypothesis*
1.00000 1.00000 1.00000
\end{verbatim}

Finally the file {\tt BCT\_estimates.txt} contains the posterior medians, and the lower the upper bound of the 95\% credible intervals of the measures of association, the intercepts and regression coefficients, and the standard deviations under  the unconstrained model. For this analysis the file looks as follows.

\begin{verbatim}
Estimates were obtained under the unconstrained model
 
Correlation matrix
 
Population  1
 
lower bound of 95%-CI
  1.000
  0.459   1.000
 -0.187  -0.392   1.000
 
median
  1.000
  0.712   1.000
  0.094  -0.060   1.000
 
upper bound of 95%-CI
  1.000
  0.865   1.000
  0.358   0.274   1.000
 
 
B-matrix with intercepts and regression coefficients

Population  1
 
lower bound of 95%-CI
 -0.250  -0.361   0.763
 
median
   0.017  -0.016   1.192
 
upper bound of 95%-CI
  0.280   0.330   1.655
 
 
standard deviations
 
Population  1
 
lower bound of 95%-CI
  0.791   1.000   1.000
 
median
  0.938   1.000   1.000
 
upper bound of 95%-CI
  1.140   1.000   1.000
\end{verbatim}
For example the 95\% credibility interval of the association between the second and first dependent variable conditional on the covariates equals $(.459,.865)$.

\bibliographystyle{apacite}
\bibliography{refs_mulder}

\end{document}